\documentclass[11pt,english]{article}
\usepackage[utf8]{inputenc}
\usepackage{amsfonts,a4wide,cite,subfigure,babel}

\usepackage{graphicx}       
\usepackage{wrapfig}
\usepackage{amsmath}        
\usepackage{amssymb}
\usepackage{bm}

\usepackage[T1]{fontenc}
\numberwithin{equation}{section}
\usepackage{appendix}

\usepackage{hyperref}
\usepackage[usenames,dvipsnames,svgnames,table,rgb]{xcolor}
\usepackage{csquotes} 

\usepackage{epigraph}
\usepackage{geometry} 
\usepackage{setspace} 
\title{\textbf{
Do  massive neutrino states really exist? 
}}

\author{Danil Shelkovkin$^1$, Oleg V. Teryaev$^2$
    \\
    \textit{
    \small{$^{1,2}$\ Bogoliubov Laboratory of Theoretical Physics,} 
    \small{Joint Institute for Nuclear Research,}} \\
    \textit{
    \small{141980 Dubna, Moscow region, Russia}
    }
    }
    
\date{}

\begin{document}

\maketitle
\begin{abstract}
	In neutrino physics, while massive states define neutrino masses, the flavor states participating in weak interactions are governed by an off-diagonal mass matrix.
	This work examines the complete form of this mass matrix, both for a two-flavor toy model and for the general three-flavor case under two distinct mass hierarchies. Using the Monte Carlo method, we estimate the mass matrix parameters and demonstrate how its structure governs the dependence of the interaction cross section on the mass hierarchy (normal vs inverted).
	This formalism enables the treatment of processes involving neutrino exchange through a non-diagonal propagator, corresponding to a quantum field theory description. Numerical estimates for a local charged-lepton interaction via virtual neutrino exchange yield a ratio of electron-antimuon (lepton flavor violating) to electron-positron cross sections on the order of $10^{-51}$. Furthermore, the cross section exhibits a fundamental dependence on the lightest neutrino mass, which differs drastically between the two hierarchies.
	For macroscopic processes, this propagator formalism reproduces the standard neutrino oscillation probability by operating directly with the non-diagonal mass matrix, thereby circumventing the wave-packet formalism and confirming the validity of this approach.
\end{abstract}

\epigraph{\textit{--- \ldots I didn't ask you whether you believe that ghosts 
are seen, but whether you believe that they exist.}
}{\text{F. Dostoyevsky}\\ \textit{<<Crime and Punishment>>}\\  \textit{Translated by R. Pevear\\ and L. Volokhovsky}}


\section*{Introduction}  

 Typically, the neutrino which is propagating for the long distances, is considered as a real particle and thus must exist in a mass eigen state. This is achieved through the diagonalization of the mass matrix entering the respective Lagrangian. 
However, in modern approaches, particularly  in Naumov's~\cite{NauNau20} and Grimus'~\cite{Gri20}, neutrino emerges as a virtual particle.
This enables computing these propagators without diagonalizing the mass matrix.

This work investigates the neutrino mass matrix. Analytically, we derive an explicit expression for the mass matrix within the Chau-Keung parametrization of the PMNS matrix, obtaining it as a function of all fundamental parameters (Euler angles $\theta_{ij}$, CP-violating phase $\delta_{\text{CP}}$, and neutrino masses $m_i$).

Numerical estimates of the mass matrix were obtained using the Monte Carlo method, incorporating data from the global fit NuFIT 6.0~\cite{EstGonMal24} and Joint neutrino oscillation analysis from  T2K and NOvA experiments published in~\cite{T2KNOvA2025}. The study of the mass matrix dependence reveals indications of fine-tuning, which could serve as a criterion for selecting a specific neutrino mass hierarchy.

Furthermore, the work extends the Quantum Field Theory (QFT) approach to lepton number violating processes through an analysis of a two-flavor toy model of Dirac neutrinos. Our complementary approach utilizes a non-diagonal matrix propagator for neutrino flavor states, which serves to validate the formalism based on the non-diagonal mass matrix. A similar methodology was employed by M. Dvornikov~\cite{Dvo25} in analyzing neutrino oscillations in matter, where the effective mass matrix becomes non-diagonal due to interactions with matter.

We have demonstrated that local processes involving neutrino exchange, such as $l^-_\alpha l^+_\beta \to W^+ W^-$, exhibit oscillation-sensitive contributions that are strongly suppressed. The degree of suppression depends critically on the flavors of the initial leptons. For the same-flavor case ($l^-_\alpha l^+_\alpha$), the characteristic scale of the cross section is $\mathcal{O}(10^{-11})$~fbarn, which is 11 orders smaller than the leading contribution from direct lepton annihilation obtained by Drutskoy et al.\ \cite{Drutskoy:2025}. For different flavors ($l^-_\alpha l^+_\beta$), the suppression is far more severe, with the cross section falling to $\mathcal{O}(10^{-62})$~fbarn, effectively rendering it unobservable.
The same lepton flavor-violating process was discussed in early studies; for example, the Singhal et al.~\cite{SinSiNA07} paper reports a difference of 59 orders of magnitude between cross sections for Majorana neutrinos (which include heavy neutrino states) and our Dirac neutrino approach.

For macroscopic processes, we also consider a process with a change in the flavor of the initial and final leptons via neutrino exchange; however, we employ a non-diagonal propagator. Our result for the amplitude in macroscopic diagrams with two-flavor transitions agrees with that of Kobzarev et al.\ \cite{Kob82}, despite their calculation having used diagonal propagators for massive neutrinos. Both approaches yield an amplitude that reproduces the canonical oscillation dependence
\[ P_{\text{osc}} \propto \sin^2\left( \frac{\Delta m^2 L}{4E_{\text{avg}}} \right),\]
where the oscillation phase is directly linked to the neutrino masses. This formalism bridges the foundations of QFT with experimental interpretation while simultaneously confirming consistency with established quantum mechanical (QM) results.

\section{Dirac neutrino propagator}

The Lagrangian for Dirac flavor neutrinos in a diagonalize form is given by:
\begin{equation}
    \mathcal{L}_{\nu}(x) = \bar{\nu}_{\alpha}(x) (i \hat{\partial} \cdot \delta_{\alpha \beta} - M_{\alpha \beta}) \nu_{\beta}(x),
\end{equation}
where summation is over flavor states.
The process of obtaining the propagator consists of inverting the matrix operator
\begin{equation}
    \mathbb{K}_{\alpha\beta}=i \hat{\partial} \cdot \delta_{\alpha \beta} - M_{\alpha \beta}.
\end{equation}
The propagator in momentum space, analogous to Krivoruchenko and \v{S}imkovic's in  \cite{KriSim23}, is defined as $\mathbb{K}^{-1}(p)\equiv\mathbb{D}(p)$ :
\begin{equation}
    \mathbb{D}(p) = \frac{1}{\hat{p} - \mathbb{M}}.
\end{equation}

\subsection{Two-flavor propagator}
We consider the specific case of interaction between  electron flavor  ($\alpha=e$) and muon flavor ($\alpha=\mu$), with a real symmetric mass matrix:
\begin{equation}\label{eq:Mass}
    \mathbb{M} = 
    \begin{pmatrix}
        m_e & m_{e \mu} \\
        m_{e \mu} & m_{\mu}
    \end{pmatrix},
\end{equation}
having eigenvalues $m_k$.

The propagator in momentum space takes the form:
\begin{equation}\label{eq:Prop_Full-P}
    \mathbb{D}(p) = 
    \frac{(\hat{p} + m_e)(\hat{p} + m_\mu) - m_{e \mu}^2}{(p^2 - m_1^2)(p^2 - m_2^2)} 
    \begin{pmatrix}
        m_e & m_{e \mu} \\
        m_{e \mu} & m_{\mu}
    \end{pmatrix}.
\end{equation}
$m_1$ and $m_2$ are the neutrino masses and are expressed in terms of the mass matrix invariants as
\begin{equation}\label{eq:def_mk}
    m_k^2=\cfrac{\text{tr}^2(\mathbb{M})}{2}-\det(\mathbb{M})+(-1)^k\ \cfrac{\text{tr}(\mathbb{M})}{2}\sqrt{\text{tr}^2(\mathbb{M})-4\det(\mathbb{M})},  
		\quad
		k=1,  2.
\end{equation}
It is convenient to represent it as
\begin{equation}\label{eq:Prop_Simp-P}
    \mathbb{D}(p) = \frac{(\hat{p}+\mathbb{M}) \cdot \widetilde{\mathbb{A}}(p^2) }{(p^2 - m_1^2)(p^2 - m_2^2)},
\end{equation}
where matrix $\widetilde{\mathbb{A}}(p^2)$ is
	\begin{equation}\label{eq:Define_A_tilda}
		\widetilde{\mathbb{A}}(p^2)=
		\begin{pmatrix}
			p^2+\det(\mathbb{M})-\text{tr}(\mathbb{M})\cdot m_{\mu}& \text{tr}(\mathbb{M})\cdot m_{e\mu}\\
			\text{tr}(\mathbb{M})\cdot m_{e\mu} & p^2+\det(\mathbb{M})-\text{tr}(\mathbb{M})\cdot m_{e}\\
		\end{pmatrix}.
	\end{equation}

If $m_{e\mu}$ tends to 0 then 
\begin{equation}
	\widetilde{\mathbb{A}}(p^2)=
	\begin{pmatrix}
		p^2- m_{\mu}^2& 0\\
		0 & p^2- m_{e}^2\\
	\end{pmatrix}
\end{equation}
and  propagator \eqref{eq:Prop_Simp-P} transform to diagonal form
\begin{equation}
	\lim\limits_{m_{e\mu}\to 0}\mathbb{D}(p) = \text{diag}\left(
	\frac{\hat{p}+m_e}{p^2 - m_e^2},\quad \frac{\hat{p}+m_{\mu}}{p^2 - m_{\mu}^2}
	\right)
\end{equation}

\subsection{Propagator and mixing angle}
It is interesting to express the resulting propagator \eqref{eq:Prop_Simp-P} in terms of mixing angle and neutrino masses.
	\begin{equation}
		\mathbb{M}=U\ \text{diag}(m_1,  m_2)\ U^{\dagger},  
	\end{equation}
	where
	$m_1,\ m_2 $ neutrino mass, and unitary rotation matrix is  
	\begin{equation}
		U=
		\begin{pmatrix}
			\cos\theta&\sin\theta\\
			-\sin\theta&\cos\theta\\
		\end{pmatrix}.
	\end{equation}
Expressing the components of the propagator through these quantities, we obtain
\begin{equation}\label{eq:Angle_mass}
    \begin{pmatrix}
        m_e&m_{e\mu}\\
        m_{e\mu}&m_{\mu}\\
    \end{pmatrix}=
    \begin{pmatrix}
m_1 \cos^2\theta + m_2 \sin^2\theta & (m_2 - m_1) \cfrac{\sin(2\theta)}{2} \\
(m_2 - m_1) \cfrac{\sin(2\theta)}{2} & m_1 \sin^2\theta + m_2 \cos^2\theta 
\end{pmatrix}
\end{equation}
and 
\begin{equation}\label{eq:A_the_best}
		\widetilde{\mathbb{A}}(p^2)=
		\begin{pmatrix}
			(p^2-m_1^2)\sin^2\theta+(p^2-m_2^2)\cos^2\theta&
			\cfrac{\Delta m^2}{2}\sin(2\theta)\\
			\cfrac{\Delta m^2}{2}\sin(2\theta) &
			(p^2-m_1^2)\cos^2\theta+(p^2-m_2^2)\sin^2\theta\\
		\end{pmatrix},
	\end{equation}
    where $\Delta m^2\equiv m_2^2-m_1^2$.
    
As a result of substituting \eqref{eq:Angle_mass} and \eqref{eq:A_the_best} into the \eqref{eq:Prop_Simp-P}, we obtain the full form of the propagator in terms of angles and neutrino masses
\begin{equation}
		\mathbb{D}(p)=
		\begin{pmatrix}
			D_1(p)\cdot\cos^2\theta+
			D_2(p)\cdot\sin^2\theta&
			\left[D_2(p)-D_1(p) \right] \cdot	
			\cfrac{\sin(2\theta)}{2}\\
			\left[D_2(p)-D_1(p) \right] \cdot	
			\cfrac{\sin(2\theta)}{2}&
			D_1(p)\cdot\sin^2\theta+
			D_2(p)\cdot\cos^2\theta\\
		\end{pmatrix},
	\end{equation}
and  $D_1,\ D_2$ correspond to mass neutrino propagator
\begin{equation}
		D_1(p)=\cfrac{\hat{p}+m_1}{p^2-m_1^2},\quad\quad
		D_2(p)=\cfrac{\hat{p}+m_2}{p^2-m_2^2}.
	\end{equation}

\section{Three flavor mass matrix}
Analyzing the  propagator for  Dirac neutrino with three flavors in its off-diagonal form presents several complications. 
However, it is useful to trace the connection between the mass matrix and the mixing angles for three flavors. 
When mass matrix have only real elements, we have 6 independent variables that can be expressed through 6 parameters: 3 masses  neutrino and 3 mixing angles.
But we have a Hermitian mass matrix with nine different variables and  \eqref{eq:M_flavor} is expressed through the  only seven independent parameters: 3 masses, 3 mixing angles and CP-phase $\delta$. We expect there are two  relation between mass matrix elements.
In the general case, the mass matrix is Hermitian:
\begin{equation}
\mathbb{M}_{\text{flavor}}=
\begin{pmatrix}
m_e & m_{e\mu} & m_{e\tau} \\
m_{\mu e} & m_{\mu} & m_{\mu\tau} \\
m_{\tau e} & m_{\tau\mu} & m_{\tau}
\end{pmatrix}.
\end{equation}

To represent it in the form
\begin{equation}\label{eq:M_flavor}
\mathbb{M}_{\text{flavor}} = V \cdot \text{diag}(m_1, m_2, m_3) \cdot V^\dagger,
\end{equation}

\subsection{Mass matrix in Chau-Keung parametrization}

Standard Chau-Keung parametrization\cite{ChaKeu84}  of Pontecorvo-Maki-Nakagawa-Sakata (PMNS) matrix $V_\text{PMNS}$ has three Euler angles $\theta_{12}, \theta_{23}, \theta_{13}$ and Dirac CP-phase $\delta$:
\begin{equation}
   \begin{split}
        &V_{\text{CK}}=
    \begin{pmatrix}
 c_{12} c_{13} & c_{13} s_{12} & e^{-i \delta } s_{13} \\
 -c_{23} s_{12}-c_{12} e^{i \delta } s_{13} s_{23} & c_{12} c_{23}-e^{i \delta } s_{12} s_{13} s_{23} & c_{13} s_{23} \\
 s_{12} s_{23}-c_{12} c_{23} e^{i \delta } s_{13} & -c_{12} s_{23}-c_{23} e^{i \delta } s_{12} s_{13} & c_{13} c_{23} \\
    \end{pmatrix}=\\&=
    	\begin{pmatrix}
				c_{12} & s_{12} & 0\\
				-s_{12} &c_{12} & 0\\
				0 & 0 & 1\\ 
		\end{pmatrix} 
        \begin{pmatrix}
            1&0&0\\
            0&1&0\\
            0&0&e^{i\delta}
        \end{pmatrix}
		\begin{pmatrix}
				c_{13} &0& s_{13} \\
				0&1&0\\
				-s_{13} &0&c_{13} \\
		\end{pmatrix} 
        \begin{pmatrix}
            1&0&0\\
            0&1&0\\
            0&0&e^{-i\delta}
        \end{pmatrix}
        \begin{pmatrix}
				1&0&0\\
				0&c_{23} & s_{23} \\
				0&-s_{23} &c_{23} \\ 
			\end{pmatrix}
   \end{split}
\end{equation}
We use the following notation:
\begin{equation}
	s_{ij}=\sin(\theta_{ij}),\quad
	c_{ij}=\cos(\theta_{ij}),\quad
	s2_{ij}=\sin(2\theta_{ij}),\quad
	s_{\delta}=\sin(\delta),\quad
	c_{\delta}=\cos(\delta),
\end{equation}
where $i,j\in\{1,2,3\}$.

The complete form of mass matrix \eqref{eq:M_flavor} can be expressed as a sum of Hermitian matrices:
\begin{equation}
\begin{split}
		\mathbb{M}_{\text{flavor}}&=
	\cfrac{m_1}{2}
	\cdot
	\begin{pmatrix}
		2 c_{12}^2 c_{13}^2&
		\begin{split}
			- s2_{12}c_{13} c_{23}- \\
			-e^{-i\delta}\ s2_{13} s_{23}c_{12}^2
		\end{split} &
		\begin{split}
			 s2_{12} s_{23}c_{13}- \\
			-e^{-i\delta}\ s2_{13} c_{12}^2  c_{23}
		\end{split} \\
		\begin{split}\\
			-  s2_{12}c_{13} c_{23}- \\
			-e^{i\delta}\ s2_{13} s_{23}c_{12}^2)
		\end{split} &
		\begin{split}\\
			2 s_{13}^2 s_{23}^2 c_{12}^2+ 2 s_{12}^2 c_{23}^2+ \\
			+	c_{\delta} s2_{12} s2_{23} s_{13}  
		\end{split}&
		\begin{split}\\
			s_{13}^2 s2_{23} c_{12}^2 -
			s_{12}^2 s2_{23}  +\\
			+s2_{12}s_{13}(2c_{\delta} c_{23}^2-e^{i\delta})
		\end{split} \\
		\begin{split}\\
			 s2_{12} s_{23}c_{13} -\\
			-e^{i\delta}\ s2_{13} c_{12}^2  c_{23}
		\end{split}  &
		\begin{split}\\
			s_{13}^2 s2_{23} c_{12}^2 -
			s_{12}^2 s2_{23}  +\\
			+s2_{12}s_{13}(2c_{\delta} c_{23}^2-e^{-i\delta})
		\end{split} &
		\begin{split}\\
			2 s_{13}^2 c_{12}^2 c_{23}^2+
			2 s_{12}^2 s_{23}^2-\\
			-c_{\delta} s2_{12} s2_{23} s_{13}  
		\end{split}\\
	\end{pmatrix}
	+\\
	&+\cfrac{m_2}{2}
	\cdot
	\begin{pmatrix}
		2 s_{12}^2 c_{13}^2&
		\begin{split}
			s2_{12}c_{13}c_{23}-\\
			-e^{-i\delta}s_{12}^2 s2_{13} s_{23}
		\end{split}&
		\begin{split}
			-s2_{12}c_{13}s_{23}-\\
			-e^{-i\delta} s_{12}^2 s2_{13} c_{23}
		\end{split}\\
		\begin{split}\\
			s2_{12}c_{13}c_{23}-\\
			-e^{i\delta}s_{12}^2 s2_{13} s_{23}
		\end{split}&
		\begin{split}\\
			2 s_{13}^2 s_{23}^2 s_{12}^2+ 2 c_{12}^2 c_{23}^2-\\
			-c_{\delta} s2_{12} s_{13} s2_{23}
		\end{split}&
		\begin{split}\\
			s_{12}^2 s_{13}^2 s2_{23}  -
			c_{12}^2 s2_{23}  -\\
			-s2_{12}s_{13}(2c_{\delta} c_{23}^2-e^{i\delta})
		\end{split}\\
		\begin{split}\\
			-s2_{12}c_{13}s_{23}-\\
			-e^{i\delta} s_{12}^2 s2_{13} c_{23}
		\end{split}&
		\begin{split}\\
			s_{12}^2 s_{13}^2 s2_{23}  -
			c_{12}^2 s2_{23}  -\\
			-s2_{12}s_{13}(2c_{\delta} c_{23}^2-e^{-i\delta})
		\end{split}&
		\begin{split}\\
			2 s_{12}^2 s_{13}^2  c_{23}^2+
			2 c_{12}^2 s_{23}^2+\\
			+c_{\delta} s2_{12} s2_{23} s_{13}  
		\end{split}
	\end{pmatrix}
	+\\
	&+\cfrac{m_3}{2}
	\cdot
	\begin{pmatrix}
		2 s_{13}^2&
		e^{-i\delta}s2_{13}s_{23}&
		e^{-i\delta}s2_{13}c_{23}\\
		e^{i\delta} s2_{13}  s_{23} &
		2 c_{13}^2 s_{23}^2&
		c_{13}^2 s2_{23} \\
		e^{i\delta}s2_{13}c_{23}&
		c_{13}^2 s2_{23}&
		2 c_{13}^2 c_{23}^2
	\end{pmatrix}
\end{split}
\end{equation}

\subsection{Numerical estimates}
Neutrino mass measurements do not yield absolute mass values but squared mass differences. The absolute mass scale is set by the lightest neutrino mass in a given mass ordering.

\begin{figure}[h!tbp]
	\centering
	\subfigure
	{\label{fig:MandlstamPlane}
		\includegraphics[scale=0.5]{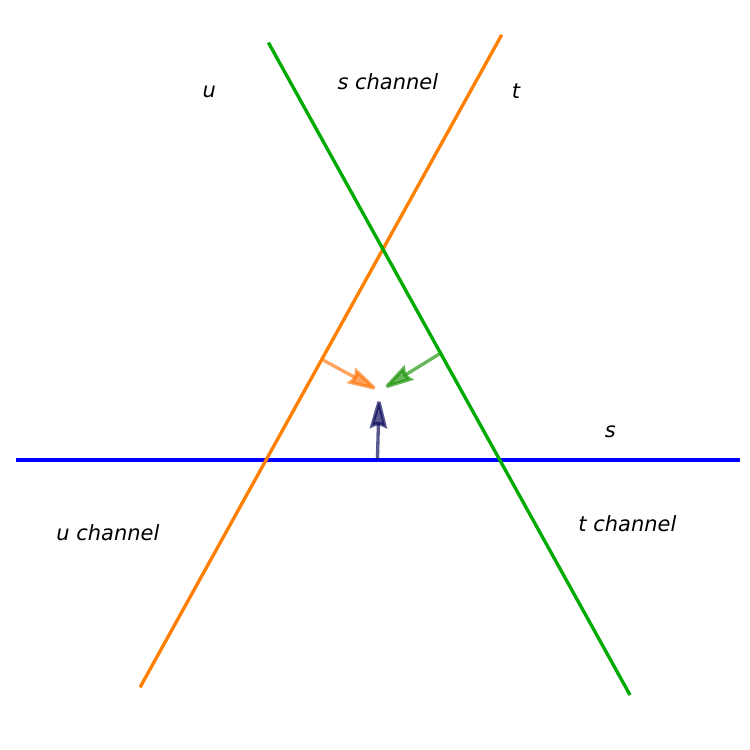}}
	\hspace{1cm}
	\subfigure
	{\label{fig:NeutrinoPlane}
		\includegraphics[scale=0.4]{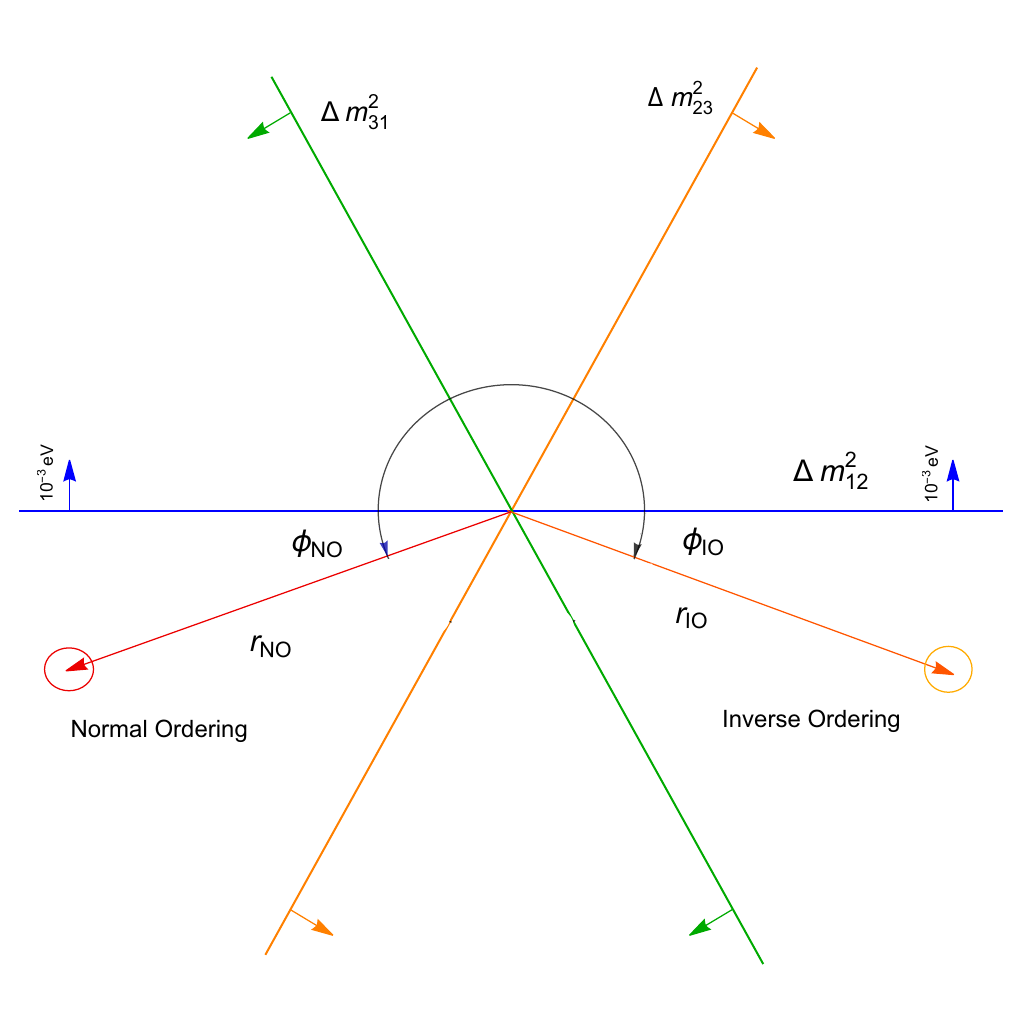}}
	\protect 
	\caption{analogue of the Mandelstam plane}
\end{figure}

It is noteworthy that the sum of the squared mass differences $\Delta m_{12}^2 + \Delta m_{23}^2 + \Delta m_{31}^2 = 0$ is analogous to the Mandelstam variables satisfying $s + t + u = \sum\limits_i m^2_i$.
Consequently, the mass eigenvalues can be represented graphically \ref{fig:NeutrinoPlane} as a Mandelstam plane, with the area of the central triangle degenerating to zero.
In other words, the mass hierarchy can be described by a single angle $\phi$ and a radius $r$.

For numerical estimates, we employ the standard Chau-Keung parametrization\cite{ChaKeu84}, where the mixing angles and the CP-violating phase are experimentally measured quantities. The current best-fit values for these parameters are provided by the global analysis of NuFIT v6.0\cite{EstGonMal24}, as well as by a recent Joint neutrino oscillation analysis from the T2K and NOvA experiments published in Nature\cite{T2KNOvA2025}.
We have 2 different situation for Inverted and Normal Ordering mass states.
\begin{table}[htbp]
		\centering
		\begin{tabular}{|l|c|c|}
			\hline
			\textbf{Parameter} & \textbf{Normal Ordering} & \textbf{Inverted Ordering} \\
			\hline&&\\[-5pt]
			$\theta_{12}\, /^\circ$ & $33.68^{+0.73}_{-0.70}$ & $33.68^{+0.73}_{-0.70}$  \\[5pt]
			$\theta_{23}\, /^\circ$ & $48.4^{+1.7}_{-2.8}$ & $48.4^{+1.7}_{-2.8}$ \\[5pt]
			$\theta_{13}\, /^\circ$ & $8.56^{+0.11}_{-0.11}$ & $8.59^{+0.11}_{-0.11}$ \\[5pt]
			$\delta_{CP}\, /^\circ$ &  $212^{+26}_{-41}$ & $274^{+22}_{-25}$ \\[5pt]
			$\cfrac{\Delta m^2_{21}}{ 10^{-5}\,\text{eV}^2}$ &$7.49^{+0.19}_{-0.19}$ & $7.49^{+0.19}_{-0.19}$ \\[8pt]
			$\cfrac{\Delta m^2_{31}}{10^{-3}\,\text{eV}^2}$ &$+2.513^{+0.021}_{-0.019}$ & $-2.409^{+0.020}_{-0.020}$ \\ [8pt]
			$\cfrac{\Delta m^2_{32}}{10^{-3}\,\text{eV}^2}$ &$+2.438^{+0.021}_{-0.019}$ & $-2.484^{+0.020}_{-0.020}$ \\ [8pt]
			\hline
		\end{tabular}
        \caption{Parameters from \href{www.nu-fit.org/?q=node/294}{NuFIT 6.0 (2024)}\cite{EstGonMal24}} and Joint neutrino oscillation analysis (2025)\cite{T2KNOvA2025}.
	\end{table}

The mass matrix values and their uncertainties are estimated using a Monte Carlo technique. First, probability distributions for the mixing angles and mass-squared differences are constructed, accounting for the mass hierarchy—specifically, the dependence on the lightest neutrino mass, which is bounded by the cosmological upper limit. These distributions are then used to generate mass matrices as functions of the lightest neutrino mass.

The resulting distribution of the mass matrix elements yields their $1\sigma$ uncertainties. The dependence of neutrino mass matrix elements on the lightest neutrino mass for both normal and inverted hierarchies is shown in Appendix~\ref{app:A}.

 Normal hierarchy mass matrix ($\mathbb{M}_{\text{NH}}\times10^{2} \text{ eV}$), defined under the assumption  $m_1=0$
\begin{equation}\label{eq:M_NH}
	\begin{split} 
		\mathbb{M}_{\text{NH}}(m_1=0)\simeq
		\begin{pmatrix}
			0.372_{-0.011}^{+0.010} & -0.185_{-0.056}^{+0.066} & -0.752_{-0.035}^{+0.055} \\
			-0.185_{-0.056}^{+0.066} & 2.671_{-0.071}^{+0.084} & 2.152_{-0.017}^{+0.011} \\
			-0.752_{-0.035}^{+0.055} & 2.152_{-0.017}^{+0.011} & 2.837_{-0.085}^{+0.070}
		\end{pmatrix}&
		+\\
		+i
		\begin{pmatrix}
			0 & 0.464_{-0.234}^{+0.042} & 0.457_{-0.243}^{+0.045} \\
			-0.464_{-0.042}^{+0.234} & 0 & -0.057_{-0.004}^{+0.026} \\
			-0.457_{-0.045}^{+0.243} & 0.057_{-0.026}^{+0.004} & 0
		\end{pmatrix}&.
	\end{split}
\end{equation}

Ratio for the inverse hierarchy ($\mathbb{M}_{\text{IH}}\times10^{2} \text{ eV}$) with respect to the $m_3=0$ case:
\begin{equation}\label{eq:M_IH}
	\begin{split}
		\mathbb{M}_{\text{IH}}(m_3=0)\simeq
		\begin{pmatrix}
			4.821_{-0.021}^{+0.021} & -0.004_{-0.221}^{+0.243} & -0.053_{-0.211}^{+0.232} \\
			-0.004_{-0.221}^{+0.243} & 2.292_{-0.065}^{+0.080} & -2.417_{-0.012}^{+0.012} \\
			-0.053_{-0.211}^{+0.232} & -2.417_{-0.012}^{+0.012} & 2.777_{-0.079}^{+0.066}
		\end{pmatrix}&
		+\\
		+i
		\begin{pmatrix}
			0 & -0.529_{-0.015}^{+0.048} & -0.483_{-0.018}^{+0.020} \\
			0.529_{-0.048}^{+0.015} & 0 & -0.005_{-0.000}^{+0.000} \\
			0.483_{-0.020}^{+0.018} & 0.005_{-0.000}^{+0.000} & 0
		\end{pmatrix}&.
	\end{split}
\end{equation}
   
\begin{figure}[h!tbp]
  \centering
  \subfigure
  {\label{fig:NHRe12}
  \includegraphics[width=0.45\textwidth]{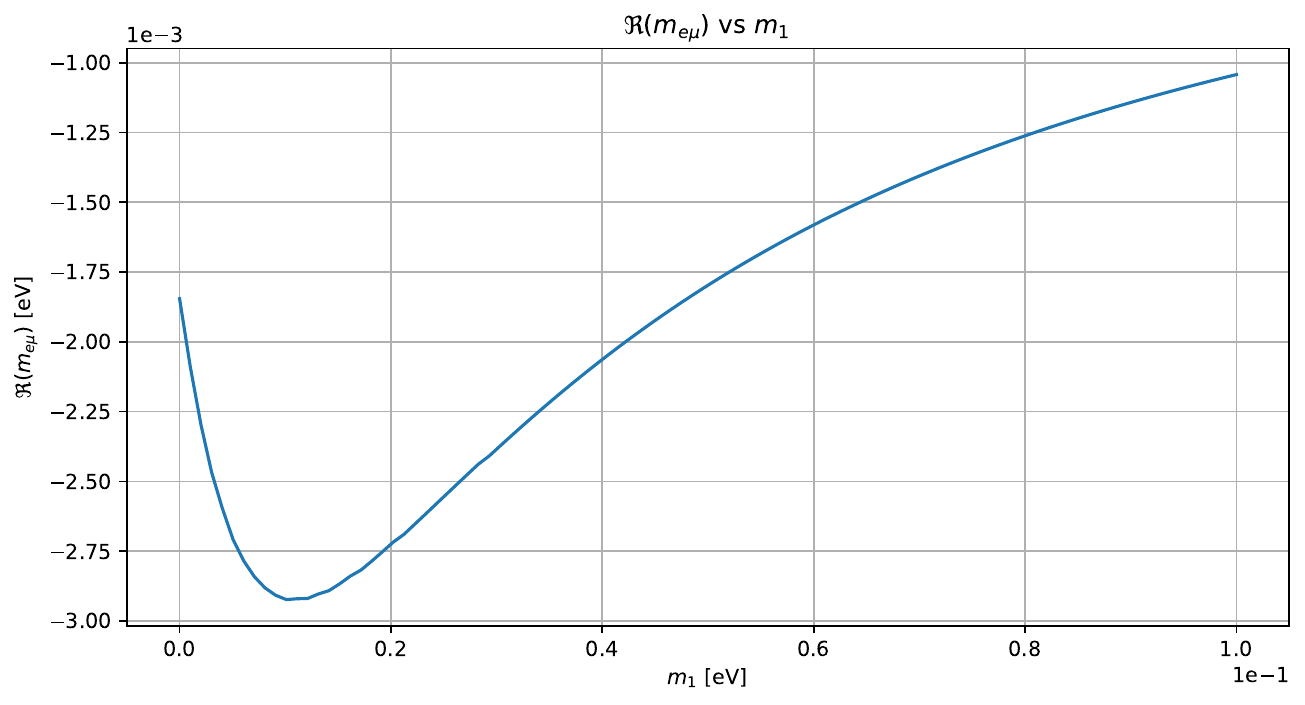}}
  \hspace{1cm}
  \subfigure
  {\label{fig:IHRe12}
  	\includegraphics[width=0.45\textwidth]{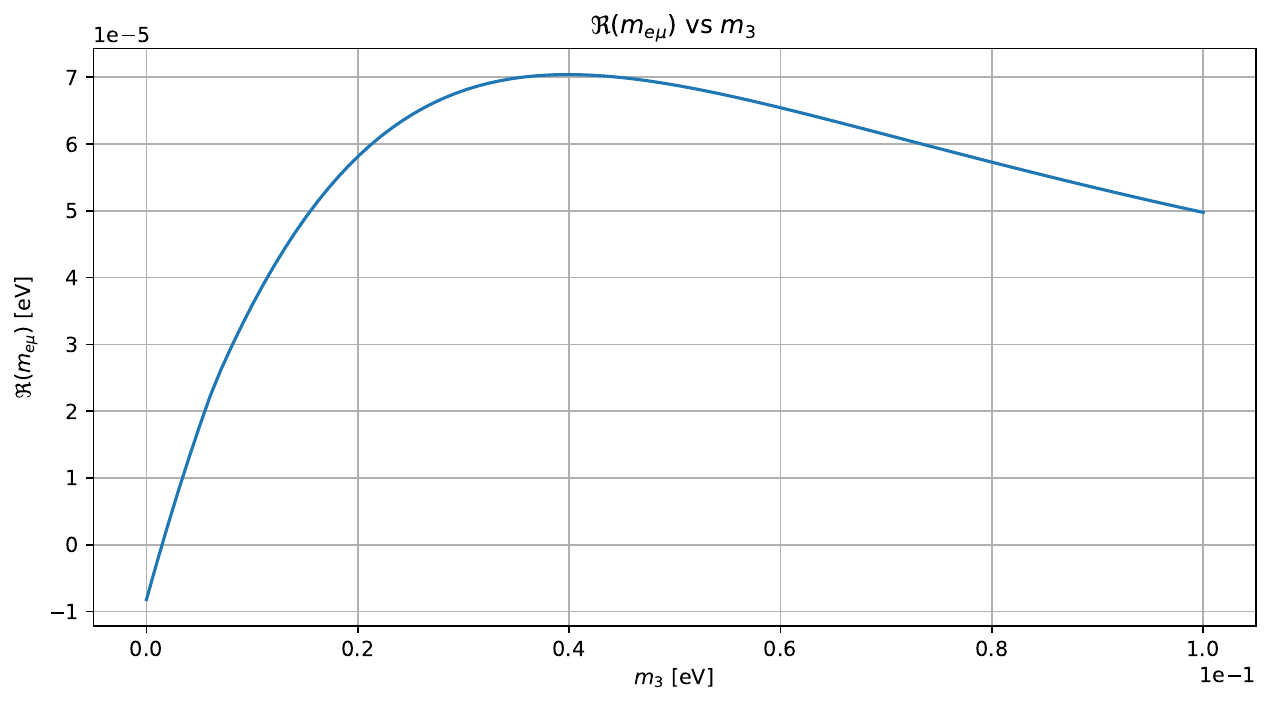}}
  \hspace{1cm}
  \protect 
\caption{Plots of $\Re(m_{e\mu})$ for both neutrino mass hierarchies: normal hierarchy on the left and inverted hierarchy on the right, as a function of the lightest neutrino mass.}
\label{kkkkk}
\end{figure}

    It was observed that for the inverted hierarchy (IH), the real values $\Re(m_{e\mu}) = \Re(m_{\mu e})$ decrease sharply as the mass $m_3$ approaches its limit:
    \begin{equation}
        \begin{split}
            &m_3=0\times10^{-2}\text{ eV}\ \sim \ \Re(m_{e\mu})=-4.07\cdot 10^{-5},\\
            &m_3=2\times10^{-2}\text{ eV}\ \sim \ \Re(m_{e\mu})=3.66\cdot 10^{-5},\\
            &m_3=4\times10^{-2}\text{ eV}\ \sim \ \Re(m_{e\mu})=5.5\cdot 10^{-5}.
        \end{split}
    \end{equation}
Analytical formula of $\Re(m_{e\mu})$ for 
        Chau-Keung parametrization:
        \begin{equation}
            \begin{split}
                \Re(m_{e\mu})=&-
            \cfrac{m_1}{2}
	           \cdot
			( s2_{12}c_{13} c_{23}+ 
			c_\delta s2_{13} s_{23}c_{12}^2)+\\
            &+
            \cfrac{m_2}{2}
	           \cdot
			(s2_{12}c_{13}c_{23}-
			c_\delta s_{12}^2 s2_{13} s_{23}   )+\\
            &+\cfrac{m_3}{2}
	           \cdot
		c_\delta s2_{13}s_{23}
            \end{split}.
        \end{equation}
    Plots for both hierarchy (Fig. \ref{kkkkk}).

As the lightest neutrino mass increases, an asymptotic approach to zero is observed for the off-diagonal elements. This behavior originates from the following relations for the normal hierarchy:
\begin{equation}
m_2=m_1\cdot\sqrt{1+\alpha_2+\alpha_2^2}, \quad
m_3=m_1\cdot\sqrt{1+\alpha_3+\alpha_3^2},
\end{equation}
where $\alpha_2=\cfrac{m_2-m_1}{2m_1}, \ \alpha_3=\cfrac{m_3-m_1}{2m_1}$.

For the inverted hierarchy:
\begin{equation}
m_1=m_3\cdot\sqrt{1+\beta_1+\beta_1^2}, \quad
m_2=m_3\cdot\sqrt{1+\beta_2+\beta_2^2},
\end{equation}
where $\beta_1=\cfrac{m_1-m_3}{2m_3}, \ \beta_2=\cfrac{m_2-m_3}{2m_3}$.
In the large-mass limit for the lightest neutrino ($m_1$ in NH, $m_3$ in IH), $\alpha_i\to 0$ and $\beta_i\to 0$. Consequently, according to \eqref{eq:M_flavor}, the flavor mass matrix reduces to the identity form for both hierarchies:
\begin{equation}
\lim_{\alpha_i\to 0}\mathbb{M}^\text{NH}_{\text{flavor}}= m_1 \cdot\mathbb{I}, \quad
\lim_{\beta_i\to 0}\mathbb{M}^{\text{IH}}_{\text{flavor}}= m_3 \cdot \mathbb{I}.
\end{equation}

All non-diagonal elements of neutrino mass matrix exhibit power-law decay towards zero without sign inversion.
However, $\Re(m_{e\mu})$ displays distinct behavior in both mass hierarchies. For the Normal Hierarchy (NH), we observe an initial local minimum followed by asymptotic convergence to zero. In the Inverted Hierarchy (IH), $\Re(m_{e\mu})$ undergoes sign reversal, reaches a global maximum beyond cosmological bounds on $m_3$, and subsequently approaches zero from above. This pronounced peak, while cosmologically nonphysical, presents a striking feature.
The inverse  hierarchy is related to some kind  of fine-tuning for matrix. This may be an additional argument in  choosing  the normal hierarchy (NH).

\section{Weak Interaction Process}
In the Standard Model, weak interactions involve only left-handed neutrinos. The interaction Lagrangian is:
\begin{equation}
    \mathcal{L}_\text{int}(x) = -\frac{g}{\sqrt{2}} \sum_\ell \overline{\ell}(x) \gamma^\mu \mathcal{P}_L \nu_\ell(x) W_\mu^-(x) - \frac{g}{\sqrt{2}} \sum_\ell \overline{\nu}_\ell(x) \gamma^\mu \mathcal{P}_L \ell(x) W_\mu^+(x),
\end{equation}
where $\mathcal{P}_L = (1 - \gamma_5)/2$ is the left-handed projector, satisfying $\mathcal{P}_L \mathcal{P}_R = 0$. In Feynman diagrams, vertices include $\gamma^\mu \mathcal{P}_L$. Consequently, in the amplitude squared, terms with constants between vertices vanish due to:
\begin{equation}\label{eq:PL_C_PR}
    \mathcal{P}_L \cdot C \cdot \gamma^\nu \mathcal{P}_L = C \cdot \mathcal{P}_L \mathcal{P}_R \gamma^\nu = 0.
\end{equation}

\subsection{Process $l_{\alpha}^- l_{\beta}^+ \to W^+ W^-$}
We consider crossing of two  charged lepton and antilepton ($e^\mp$, $\mu^\pm$) scattering into two W-bosons and evaluate the contribution of neutrino oscillations to this local process. 
\begin{figure}[h!tbp]
  \centering
  \includegraphics[scale=1]{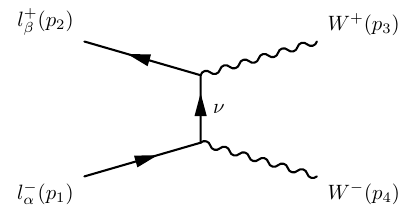}
\caption{Diagram of $l_{\alpha}^- l_{\beta}^+ \to W^+ W^-$ for t-channel.
 \label{fig:Diagrams}}
\end{figure}
For the kinematics, we assume negligible lepton masses compared to their energies and negligible neutrino masses. In the center-of-mass frame of the incoming leptons:
\begin{equation}\label{eq:Kin}
    \begin{split}
        &p_1 \simeq (|\mathbf{p}_{in}|, \mathbf{p}_{in}), \\
        &p_2 \simeq (|\mathbf{p}_{in}|, -\mathbf{p}_{in}), \\
        &p_3 \simeq p_4 \simeq (M_W, 0).
    \end{split}
\end{equation}
Given the propagator form \eqref{eq:Prop_Simp-P}, the property \eqref{eq:PL_C_PR}, and the kinematics \eqref{eq:Kin}, the amplitude squared $\mathcal{A}^2_{\alpha \beta}$ in the flavor basis simplifies to:
\begin{equation}\label{eq:M2_Cal+Kin+Simpl}
    \mathcal{A}^2 \simeq 
    \frac{8 g^4}{M_W^4} 
    \begin{pmatrix}
        M_W^4 & \text{tr}^2(\mathbb{M}) \cdot |m_{e \mu}|^2 \\
        \text{tr}^2(\mathbb{M}) \cdot |m_{e \mu}|^2 & M_W^4
    \end{pmatrix},
\end{equation}
where diagonal elements correspond to the exchange of neutrinos of the same flavor, and off-diagonal elements to oscillations.

The differential cross section for this process is:
\begin{equation}
    \frac{d \sigma}{d \Omega} = \frac{\mathcal{A}^2}{64 \pi^2 s} \simeq \frac{g^4}{16 \pi^2} \frac{1}{M_W^6}
    \begin{pmatrix}
        M_W^4 & \text{tr}^2(\mathbb{M}) \cdot |m_{e \mu}|^2 \\
        \text{tr}^2(\mathbb{M}) \cdot |m_{e \mu}|^2 & M_W^4
    \end{pmatrix}.
\end{equation}

Considering the numerical estimates for $|m_{e\mu}|^2$ from \eqref{eq:M_NH} and \eqref{eq:M_IH}, the amplitude squared \eqref{eq:M2_Cal+Kin+Simpl} will yield:
For the case of identical flavors (diagonal elements), the cross sections are suppressed by $M_W^2$ for both hierarchies and are approximately equal to $d\sigma\simeq10^{-11}$~fbarn.

Regarding the off-diagonal element, the cross sections are suppressed by more than $M_W^6$ and are approximately equal to $d\sigma\simeq10^{-62}$~fbarn for both hierarchies.

\begin{figure}[h!tbp]
	\centering
	\subfigure
	{\label{fig:Cross-a}
		\includegraphics[scale=0.32]{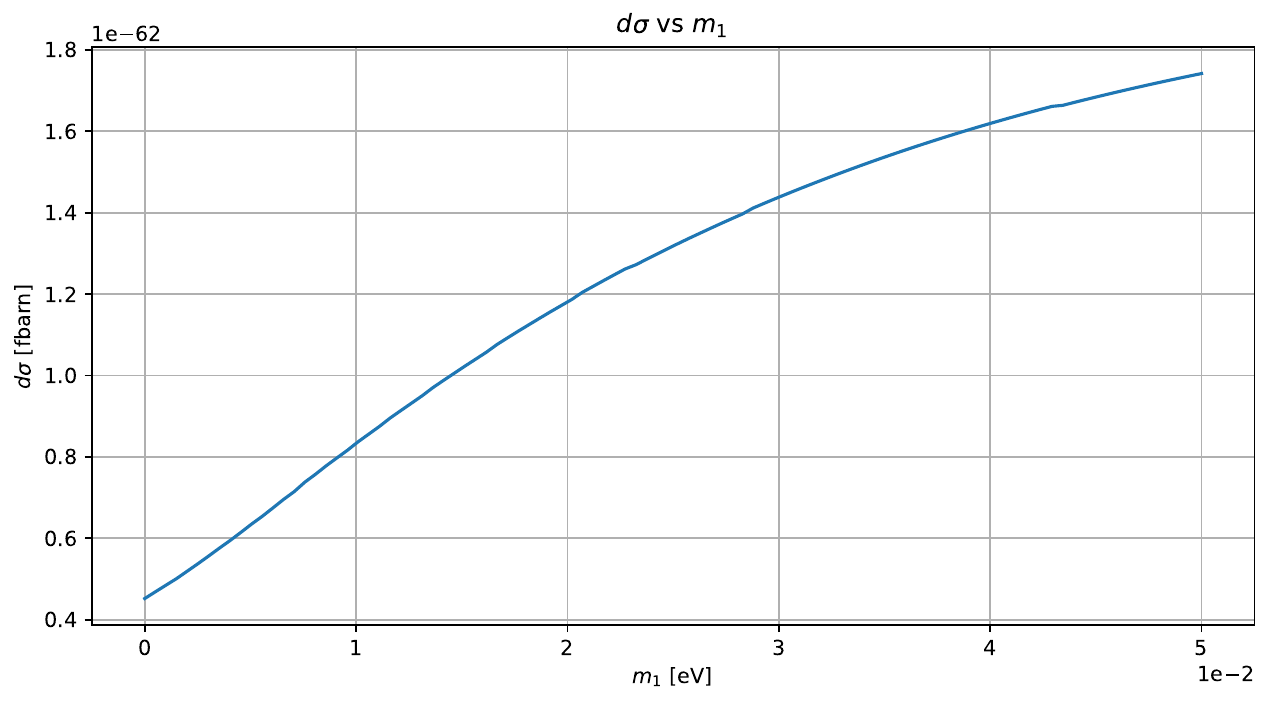}}
	\hspace{1cm}
	\subfigure
	{\label{fig:Cross-b}
		\includegraphics[scale=0.32]{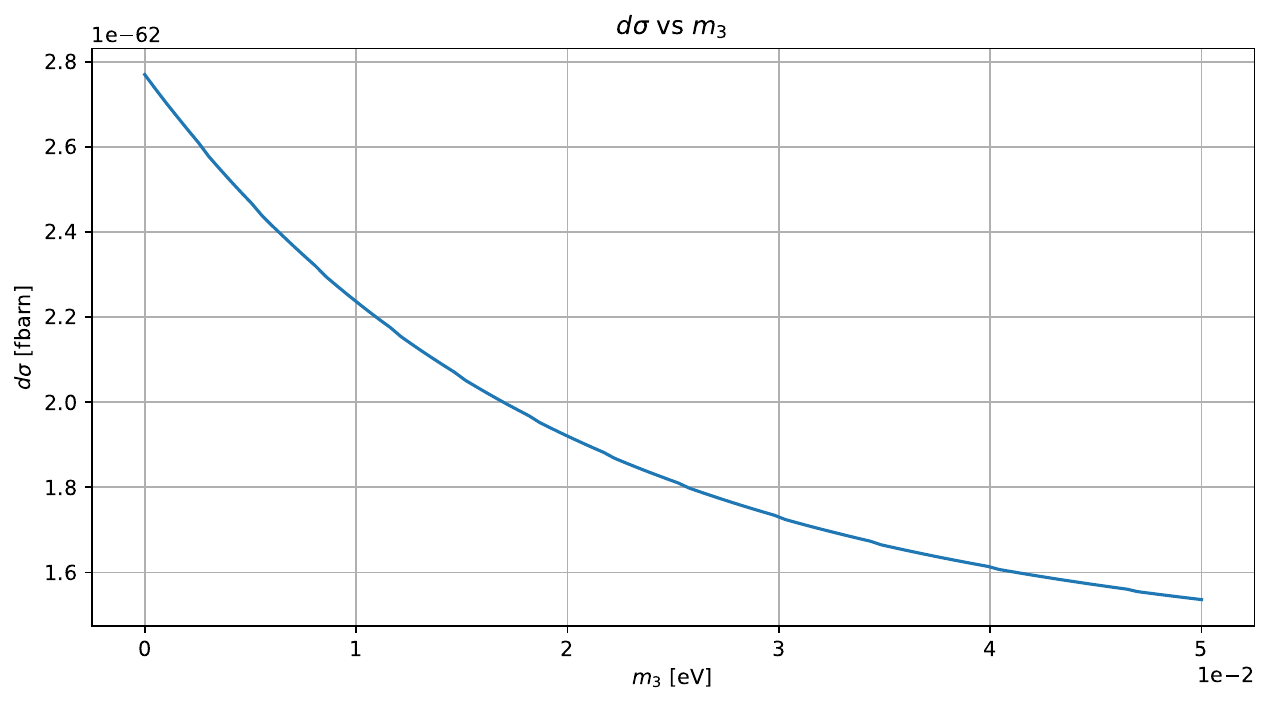}}
	\protect 
\caption{Plots of $\frac{d\sigma}{d\Omega}$ for both neutrino mass hierarchies: normal (left)  and inverted (right), as a function of the lightest neutrino mass $m_1$ and $m_3$, respectively.}
\label{fig:Cross}
\end{figure}

Numerical estimates of the cross section dependence on the increasing lightest neutrino mass show that, on the one hand, the approximate cross section increases with $m_1$ for the \textit{normal hierarchy} (fig.~\ref{fig:Cross-a}). On the other hand, for the \textit{inverted hierarchy}, this value decreases as $m_3$ approaches its maximum value (fig.~\ref{fig:Cross-b}).

It should be noted that the electron-positron annihilation process with W-boson pair production was considered in \cite{Drutskoy:2025}, where the obtained cross section is 11 orders of magnitude larger. This discrepancy arises primarily because we do not consider heavy neutrinos, and secondly, because we only examine neutrino exchange, which is significantly smaller than the cross sections for W-boson pair annihilation and their interference.

As for processes with lepton number violation, a similar reaction was studied in \cite{SinSiNA07} for Majorana neutrinos. For Dirac neutrinos, the difference in cross sections amounts to 59 orders of magnitude, which is also attributed to the consideration of heavy neutrinos.

The ratio of flavor-changing cross sections to the electron-positron cross section for the two mass hierarchies 
\begin{equation}
	\frac{d\sigma_{\mu^+e^- \to W^+W^-}}{d\sigma_{e^+e^- \to W^+W^-}}\simeq\cfrac{\text{tr}^2(\mathbb{M}) m_{e \mu}^2}{M_W^4} 
\end{equation}
corresponds to value on the order of $10^{-51}$ for both hierarchy.
Such a small value of the cross section ratio excludes the possibility of observing these processes in current and future experiments.
This result thus validates neglecting the contributions from flavor-changing neutrino exchange interactions in the analysis of $W$-boson production processes in lepton collisions.

\subsection{Neutrino Oscillations}
\begin{figure}[h!tbp]
	\centering
		\includegraphics[scale=1]{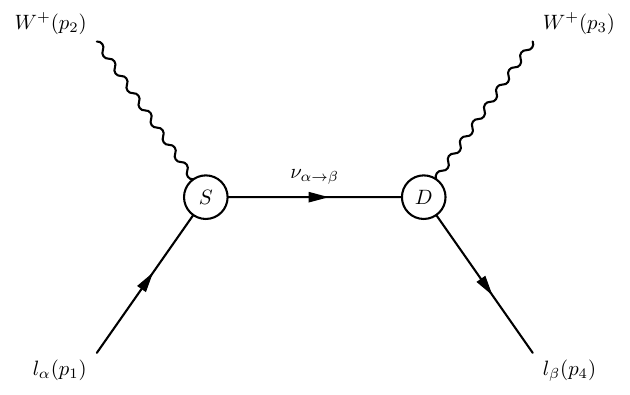}
	\caption{Diagram of $l_{\alpha} W^+  \to l_{\beta} W^+$ for s-channel.
	\label{fig:s-chanel}}
\end{figure}

The manifestation of lepton flavor-violating processes via virtual neutrino exchange appears in the s-channel of macro-diagram (fig.~\ref{fig:s-chanel}), where the spatial  $\mathbf{x} - \mathbf{y} \gg 0$ and $x_0 - y_0 > 0$ separation between the source $S$ and the detector $D$ plays a significant role in neutrino oscillations. 

For simplicity, we again turn to a toy model with two flavors, which allows us to work quite straightforwardly with matrix forms.
The neutrino propagator in such a process must be considered in coordinate space, which explicitly highlights the distance L between the detector and the source.
\begin{equation}\label{eq:propfin}
	\mathbb{D}(x-y) = 
	\frac{1}{4\pi^2 \text{L}}	\sum_{k=1,  2}
	{(-1)^k}
	\frac{\text{p}_k }{2\omega_k} 
	\left[\gamma^0\omega_k+ \mathbb{M} -\left(\text{p}_k+\cfrac{i}{\text{L}}\right) \mathbf{n}_{\text{L}}\bm{\gamma}   
	\right]
	\mathbb{A}_{k}\
	e^{i \text{p}_k \text{L}}
\end{equation}
where the momenta $p_k$ correspond to the $k$ neutrino masses  
\begin{equation}
	\text{p}_k=\sqrt{ E_{\text{avg}}^2-m_k^2}
\end{equation}
 and $E_{\text{avg}}$ denote  the average energy at the detector and source, respectively.
The matrices $\mathbb{A}_k$ are expressed through the mixing angle $\theta$, defined below in \eqref{eq:sin2theta}
\begin{equation}\label{eq:Define_A_Angle}
	\mathbb{A}_1\equiv
	\begin{pmatrix}
		-\cos^2\theta & \cfrac{\sin(2\theta)}{2}\\
		\cfrac{\sin(2\theta)}{2} & -\sin^2\theta\\
	\end{pmatrix},  \quad
	\mathbb{A}_2\equiv
	\begin{pmatrix}
		\sin^2\theta & \cfrac{\sin(2\theta)}{2}\\
		\cfrac{\sin(2\theta)}{2} & \cos^2\theta\\
	\end{pmatrix}.
\end{equation}

A propagator of similar form to \eqref{eq:propfin} was derived by Kobzarev et al.~\footnote{See formula (1.5) of Appendix 1 in~\cite{Kob82}} in a general diagonalized representation for three flavors:
\begin{equation}
	\mathbb{D}(x-y)\propto \frac{1}{4\pi \text{L}} \left[ \varepsilon \gamma_0 + m_A - \left( \text{p}_A + \cfrac{i}{\text{L}} \right) \mathbf{n}_{\text{L}}\bm{\gamma}   \right] U_{iA}^{\dagger}U_{kA} e^{i \text{p}_A \text{L}}.
\end{equation}

The specific form of the matrix propagator, consisting of a sum, leads to an amplitude of the form:
\begin{equation}
	\mathcal{A}_{\alpha \beta} \propto \frac{1}{\text{L}}\left( \mathbb{A}_{2}^{ \alpha \beta} 
	\exp \left\{ i \text{L} \sqrt{ E_{\text{avg}}^2-m_2^2} \right\} - \mathbb{A}_{1}^{ \alpha \beta} \exp \left\{ i \text{L} \sqrt{ E_{\text{avg}}^2-m_1^2} \right\} \right),
\end{equation}
the square of which produces interfering terms responsible for flavor change in the process.

Omitting detailed calculations, we find that considering processes with two flavor changes allows us to reproduce the standard formula for the probability of neutrino oscillations:
\begin{equation}\label{eq:Final_Oscill}
	P_{e\to\mu} \propto
	{\sin^2(2\theta)}\ 
	\sin^2\left(\cfrac{\Delta m^2}{4E_{\text{avg}}}\text{L}\right),
\end{equation}
where the quantities
\begin{equation}\label{eq:sin2theta}
	\Delta m^2 = \text{tr}(\mathbb{M}) \sqrt{\text{tr}^2(\mathbb{M}) - 4 \det(\mathbb{M})}\quad \text{and}\quad
	\sin(2\theta)=\cfrac{2 m_{e\mu}}{\sqrt{\text{tr}^2(\mathbb{M}) - 4 \det(\mathbb{M})}}
\end{equation}
are expressed through the invariants of the mass matrix.

\section*{Conclusions} 
The consideration of processes with lepton number violation shows that it makes no difference whether one works in the mass or flavor basis: the same expressions are obtained, although flavor states appear in weak interactions.

Working in the flavor basis, we derived the matrix propagator both in its general form and with explicit angular dependence. This may seem like an overly elaborate construction for processes that can be easily diagonalized in the mass basis. However, in situations where diagonalization is not straightforward  (for instance, in interactions with matter~\cite{Dvo25}), it becomes necessary to employ matrix propagators.

For the neutrino mass matrix, we obtained an explicit analytical expression in terms of the standard parameters: mixing angles, CP-violating phases, and neutrino masses. To our knowledge, such a representation has not been previously reported in the literature. Based on the global fit data from NuFIT~6.0~\cite{EstGonMal24} and the joint T2K and NOvA analysis~\cite{T2KNOvA2025}, we applied the Monte Carlo method to generate distributions of the mass matrix elements, thereby obtaining their uncertainties and numerical values for both normal and inverted hierarchies.

The data also allowed us to study the dependence of the mass matrix elements on the lightest neutrino mass for each hierarchy. Interestingly, in the inverted hierarchy, the real part of the element $m_{e\mu}$ is very small: its magnitude reaches a maximum at positive values and then asymptotically approaches zero. Given the large uncertainty in the estimate of this element, its precise value cannot be determined. Nevertheless, such behavior may indicate a fine-tuning that seems to flavor the normal hierarchy.

Next, we considered the crossing process of two oppositely charged leptons producing a pair of W bosons, $l_{\alpha}^- l_{\beta}^+ \to W^+ W^-$, within a two-flavor toy model. As expected, the contribution from neutrino exchange is extremely small, $d\sigma \simeq 10^{-11}~\text{fbarn}$. The cross-section for the electron-positron case  $e^+ e^-\xrightarrow{\nu } W^+ W^-$ is seven orders of magnitude smaller than that reported in~\cite{Drutskoy:2025}, since we did not include heavy neutrinos and considered only the neutrino-exchange contribution, which is suppressed relative to the dominant $e^+ e^-\xrightarrow{Z,\ \gamma } W^+ W^-$ annihilation into two W-bosons and their interference.
It should be noted that the main contribution from the neutrino exchange process will be most apparent in the interference term with the annihilation process, as it depends linearly on this matrix element.
This suggests searching for charge asymmetry in $W^+ W^-$ final states, where interference between C-odd annihilation and neutrino-exchange amplitudes could produce observable effects linear in the weak amplitude, unlike the quadratically suppressed direct contribution.

The cross-section for the lepton flavor-violating process $e^-\mu^+ \to W^-W^+$ demonstrates extreme suppression, reaching approximately $d\sigma \simeq 10^{-62}~\text{fb}$ - about 51 orders of magnitude smaller than comparable flavor-conserving processes. This pattern of extreme suppression aligns with findings in lepton number violating interactions; notably, \cite{SinSiNA07} reported a difference of 59 orders of magnitude between Majorana and Dirac neutrino cross sections in similar processes. This dramatic discrepancy can be naturally explained within frameworks incorporating heavy neutrino states, particularly through the seesaw mechanism, where the inclusion of heavy Majorana neutrinos significantly enhances the amplitude of lepton number violating processes compared to the Dirac case.

The normal and inverted hierarchies exhibit opposite behaviors: in the normal hierarchy (NH), the cross-section increases with the lightest neutrino mass $m_1$, while in the inverted hierarchy (IH), it decreases with the lightest mass $m_3$. However, such differences cannot be tested experimentally due to the extremely small cross-sections.

Flavor-changing processes become apparent when neutrinos propagate over large distances, necessitating the consideration of macroscopic (long-distance) diagrams. The calculation of such a diagram within the QFT framework reproduces the well-known quantum mechanical formula for neutrino oscillations. Furthermore, the expressions obtained for the matrix propagator agree with those in~\cite{Kob82}, while in our case all formulas are expressed explicitly in terms of the invariants of the neutrino mass matrix.

Coming back to the question posed in the title we may say that there seems to be no definite answer as soon as weak interactions are considered. The appearance of massive and flavor states may be considered as a  kind of Bohr's complementarity.  At the same time, consideration of neutrino helicity flip due to interactions with gravity (see e.g. \cite{Dvornikov:2006ji,Teryaev:2016edw,Baym:2021ksj}) may require 
the appearance of definite massive states.

\section*{Acknowledgments}

We are most indebted to V.A. Naumov for useful discussions and valuable  comments.


\newpage
\appendix
\section{Plots of mass matrix element dependence for NH and IH. }\label{app:A}

Plots of mass matrix element dependence on $m_1$ for Normal Hierarchy
\begin{figure}[h!tbp]
	\centering
	\subfigure
	{\label{fig:Re11}
		\includegraphics[width=0.45\textwidth]{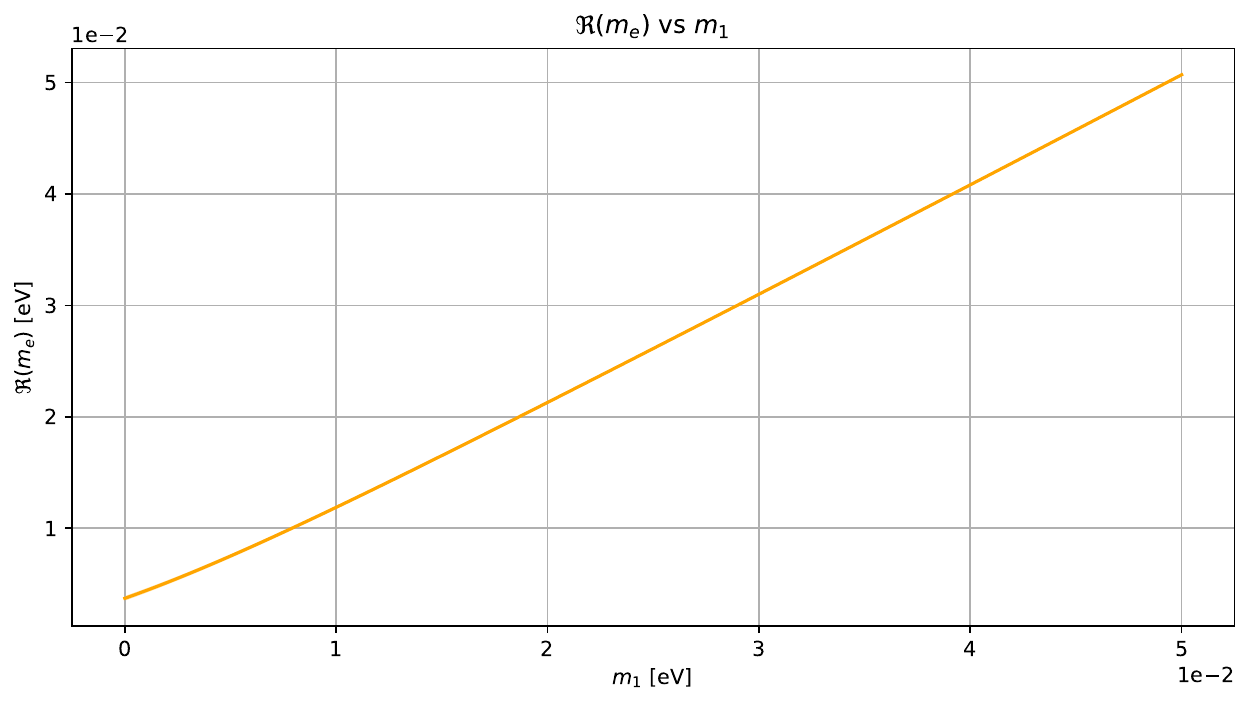}}
	\hspace{1cm}
	\subfigure
	{\label{fig:Re22}
		\includegraphics[width=0.45\textwidth]{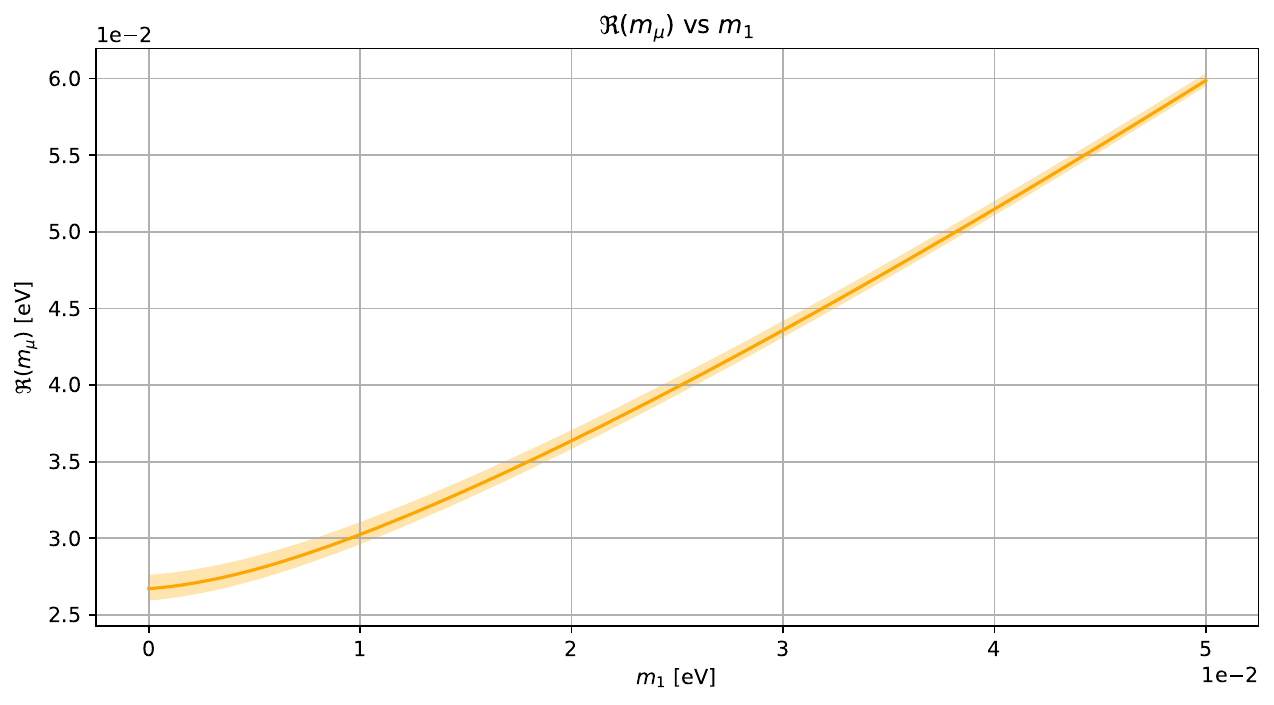}}
	\hspace{1cm}
	\subfigure
	{\label{fig:Re33}
		\includegraphics[width=0.45\textwidth]{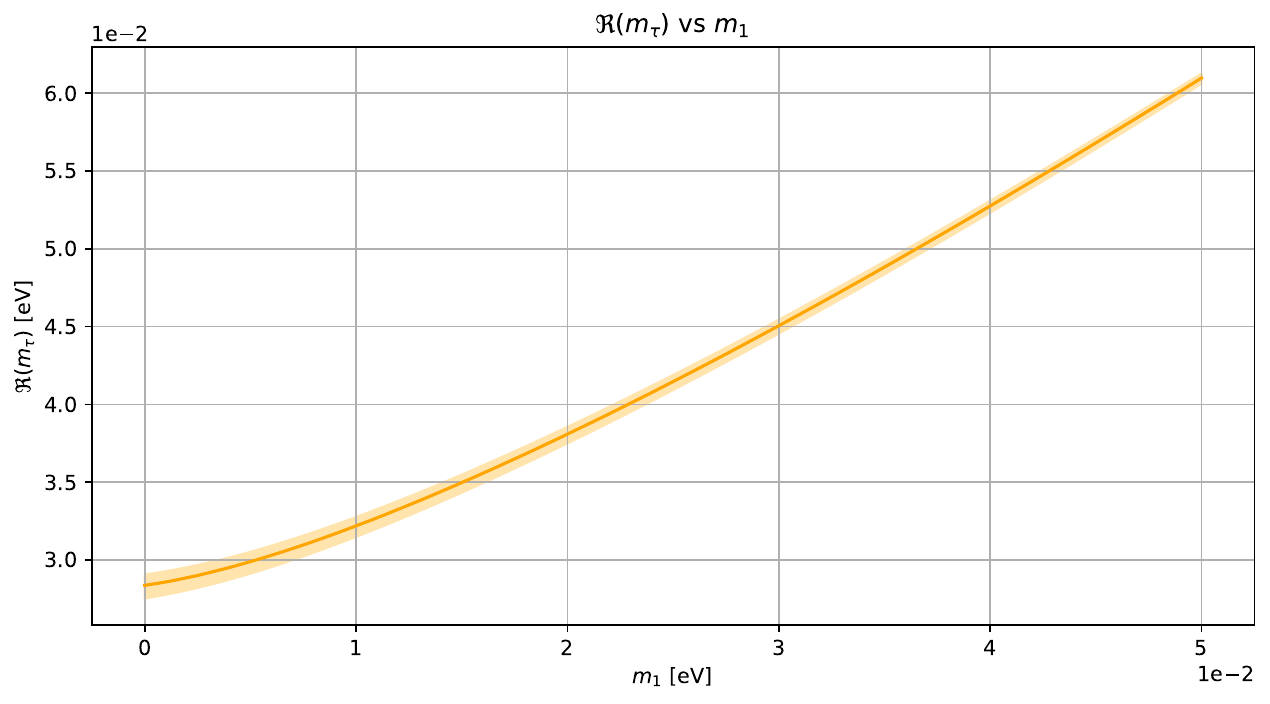}}
	\protect 
	\caption{Diagonal elements $m_{e}$, $m_{\mu}$, $m_{\tau}$ of the neutrino mass matrix in NH versus $m_1$}
\end{figure}
\begin{figure}[h!tbp]
	\centering
	\subfigure
	{\label{fig:Re12}
		\includegraphics[width=0.45\textwidth]{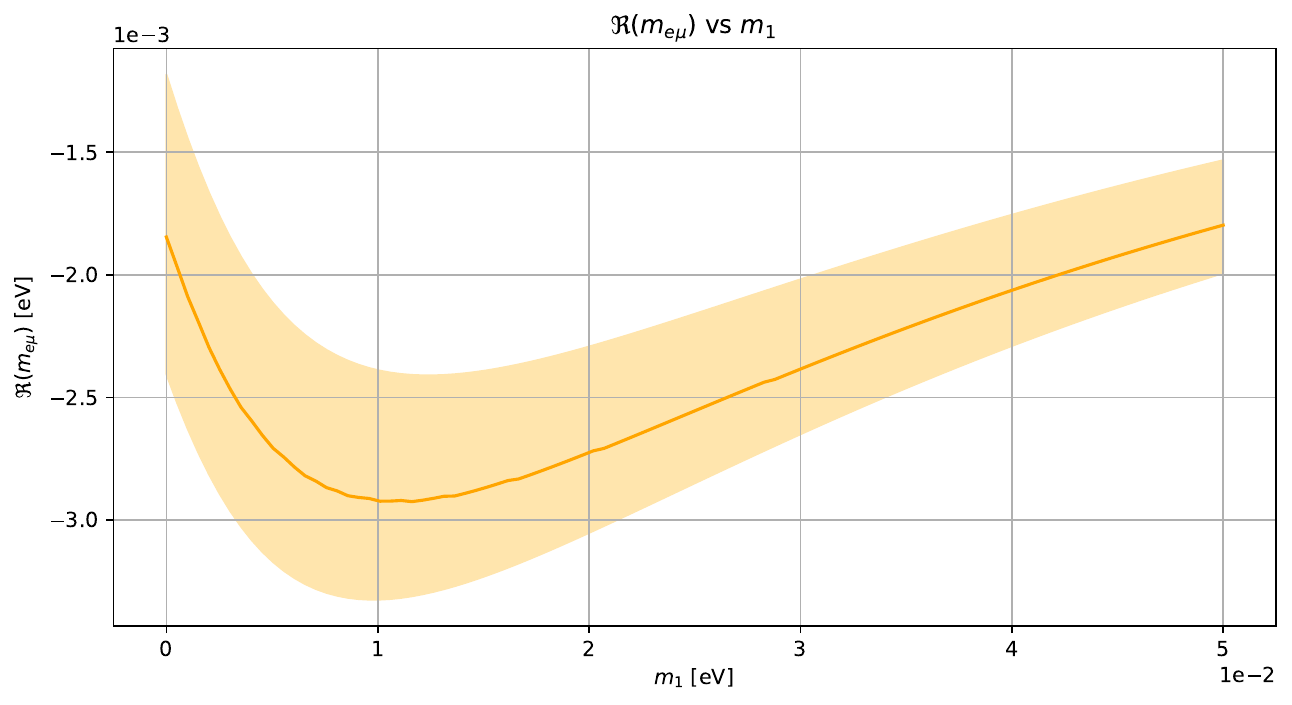}}
	\hspace{1cm}
	\subfigure
	{\label{fig:Im12}
		\includegraphics[width=0.45\textwidth]{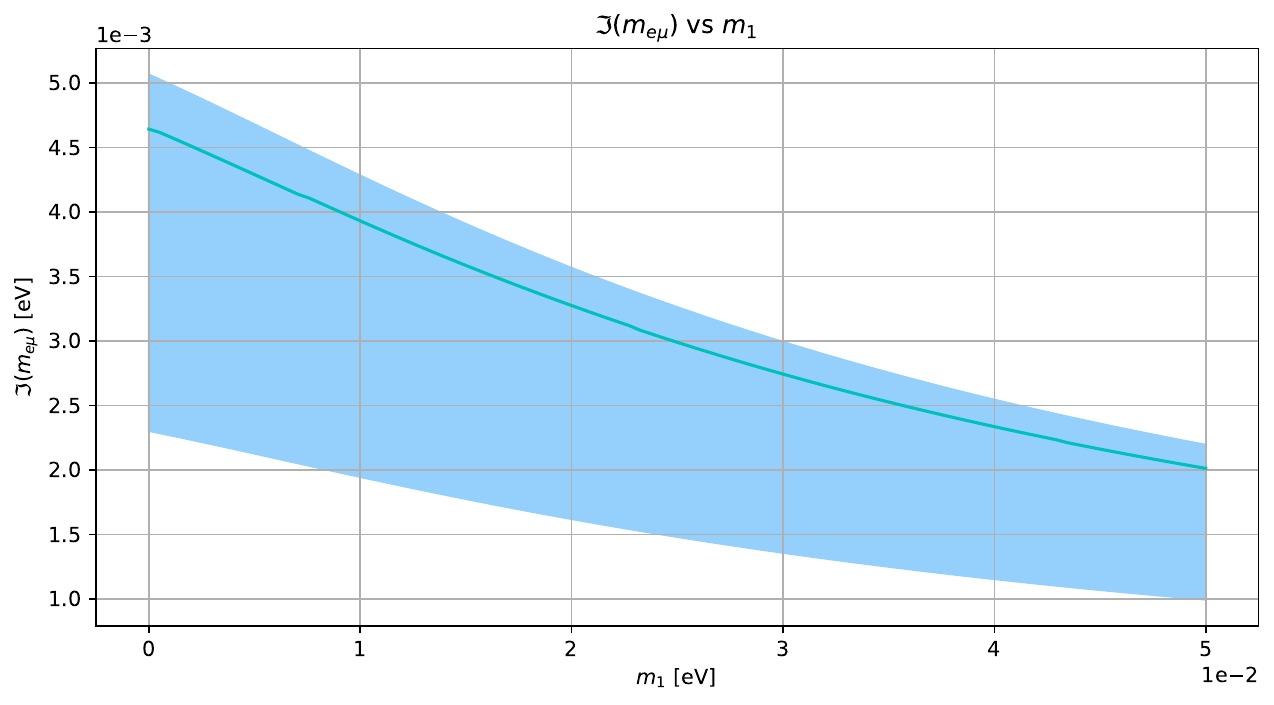}}
	\protect 
	\caption{Real and imaginary components of $m_{e\mu}$ in NH as a function of $m_1$}
\end{figure}
\begin{figure}[h!tbp]
	\centering
	\subfigure
	{\label{fig:Re13}
		\includegraphics[width=0.45\textwidth]{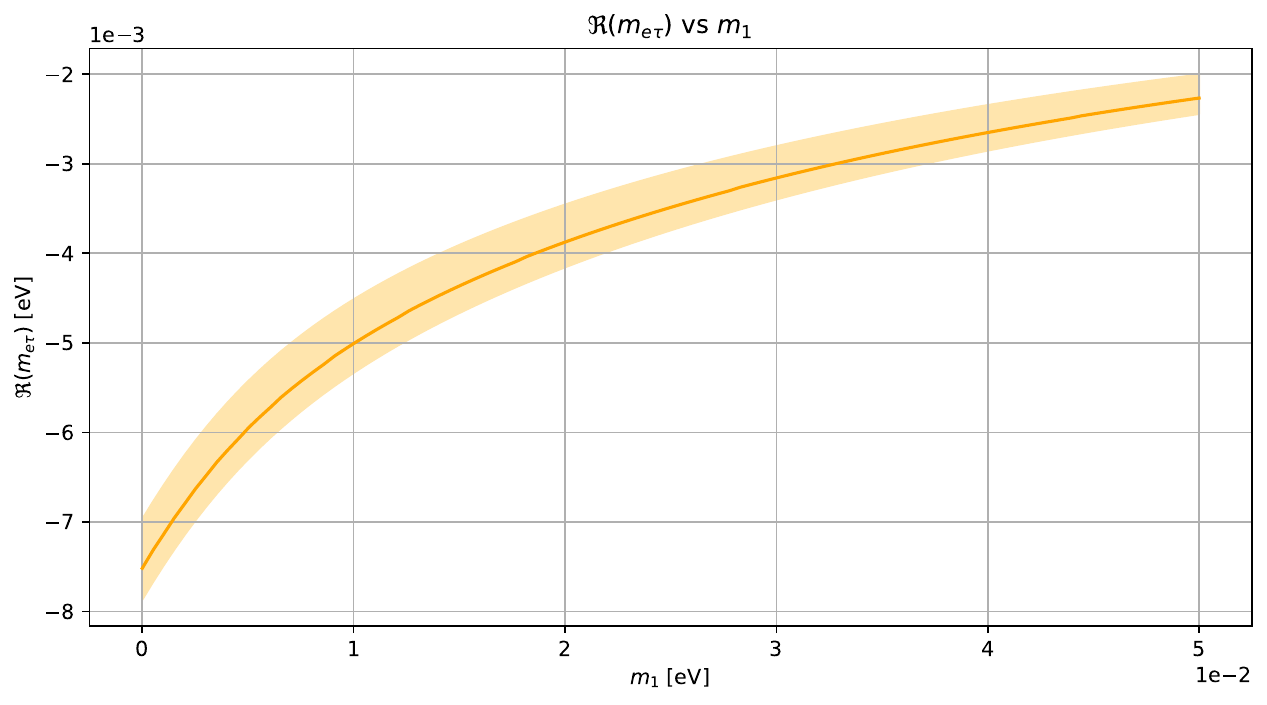}}
	\hspace{1cm}
	\subfigure
	{\label{fig:Im13}
		\includegraphics[width=0.45\textwidth]{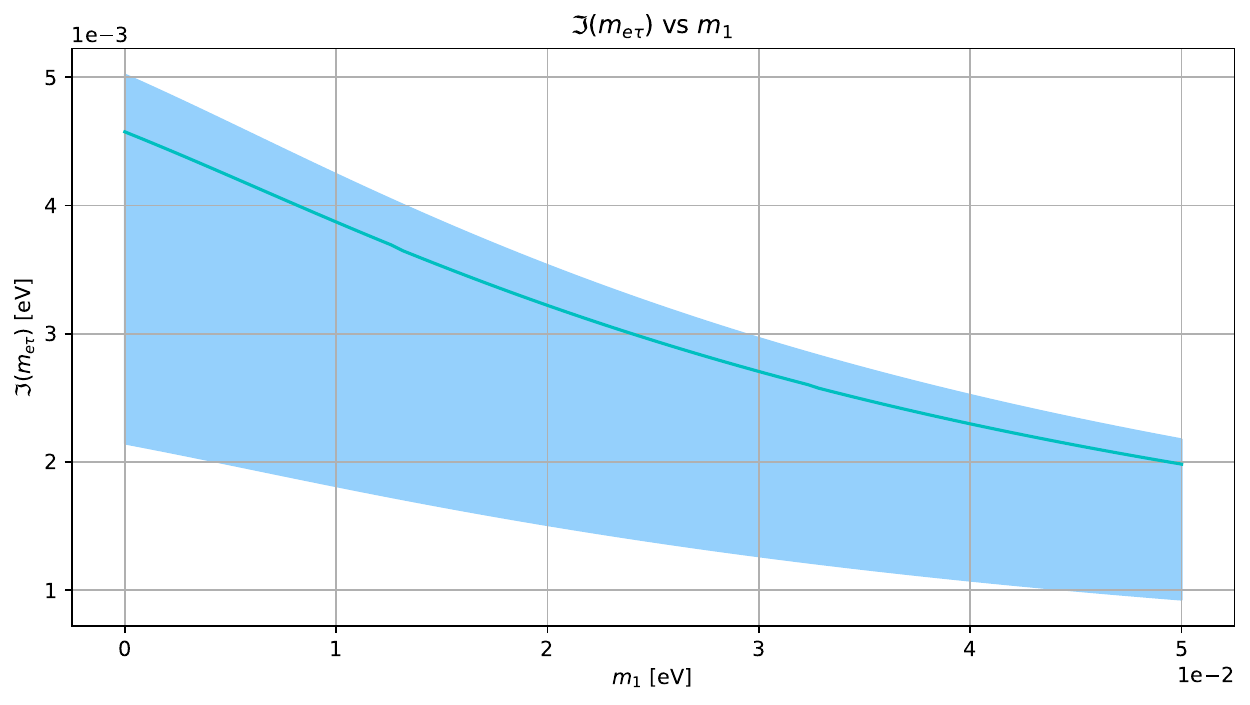}}
	\protect 
	\caption{Real and imaginary components of $m_{e\tau}$ in NH versus $m_1$}
\end{figure}
\begin{figure}[h!tbp]
	\centering
	\subfigure
	{\label{fig:Re23}
		\includegraphics[width=0.45\textwidth]{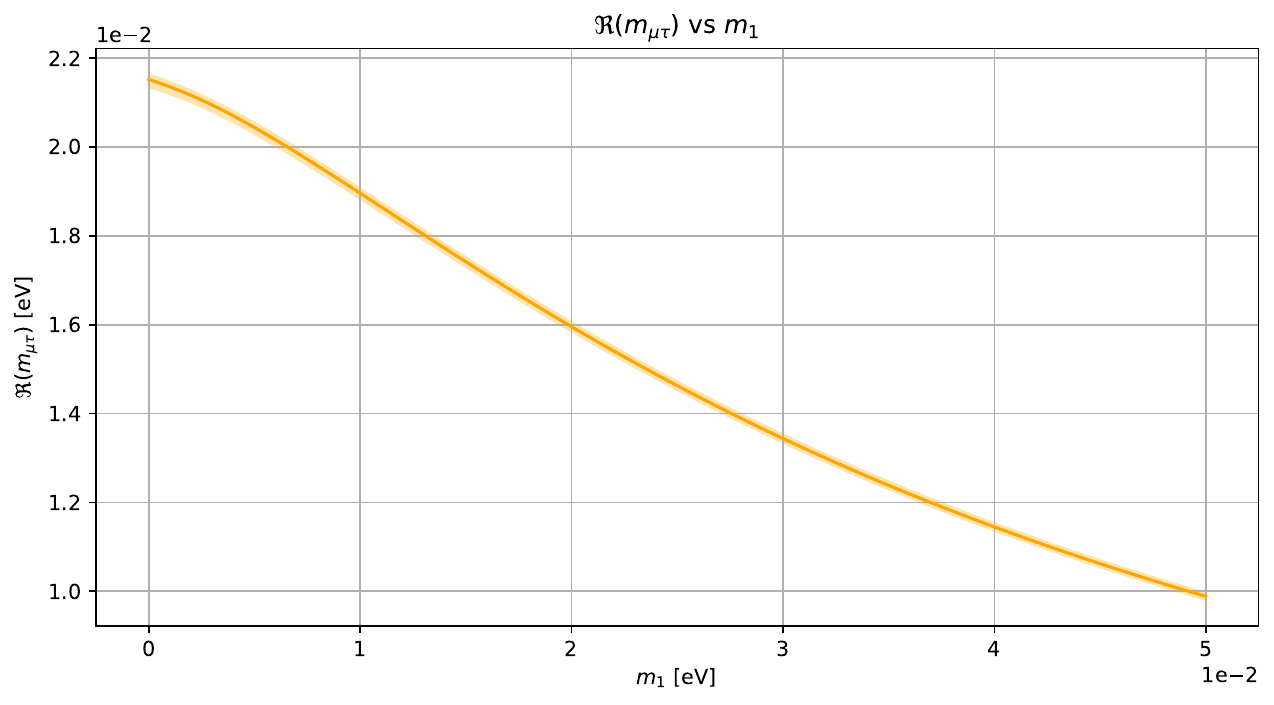}}
	\hspace{1cm}
	\subfigure
	{\label{fig:Im23}
		\includegraphics[width=0.45\textwidth]{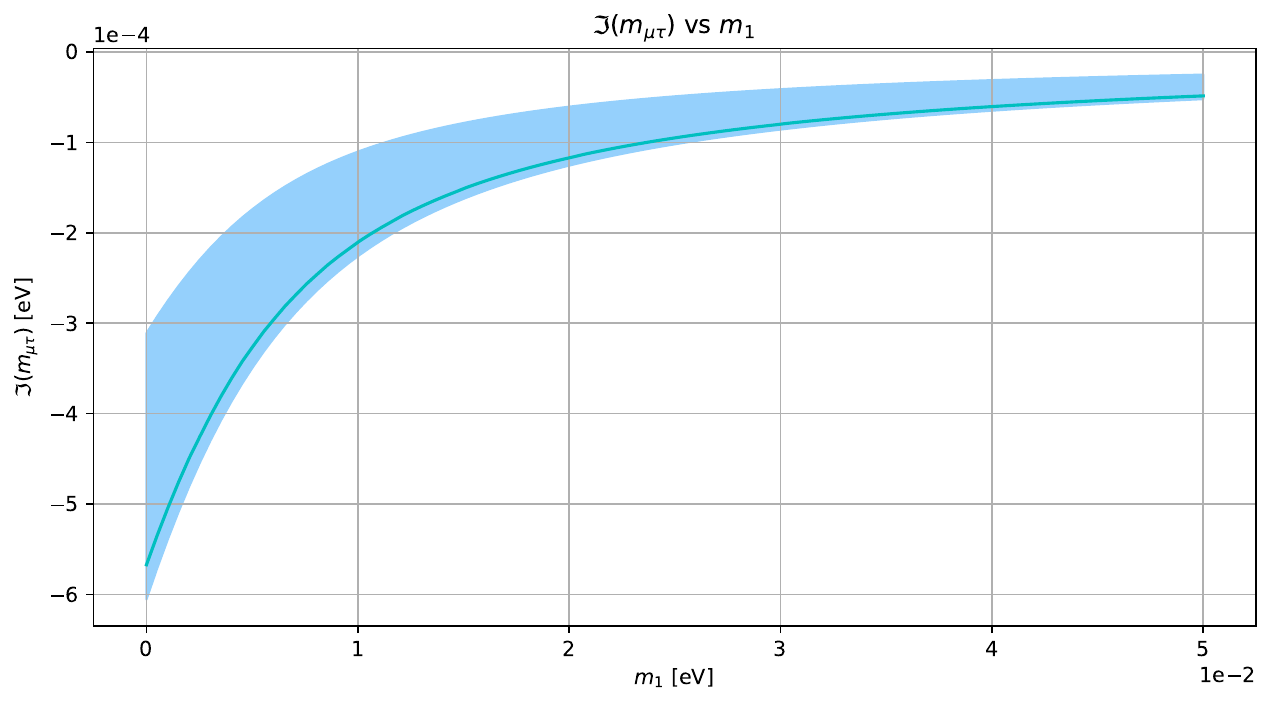}}
	\protect 
	\caption{Real and imaginary components of $m_{\mu\tau}$ in NH as a function of $m_1$}
\end{figure}
\newpage
Plots of mass matrix element dependence on $m_3$ for Inverted Hierarchy
\begin{figure}[h!tbp]
	\centering
	\subfigure
	{\label{fig:Re11_IH}
		\includegraphics[width=0.45\textwidth]{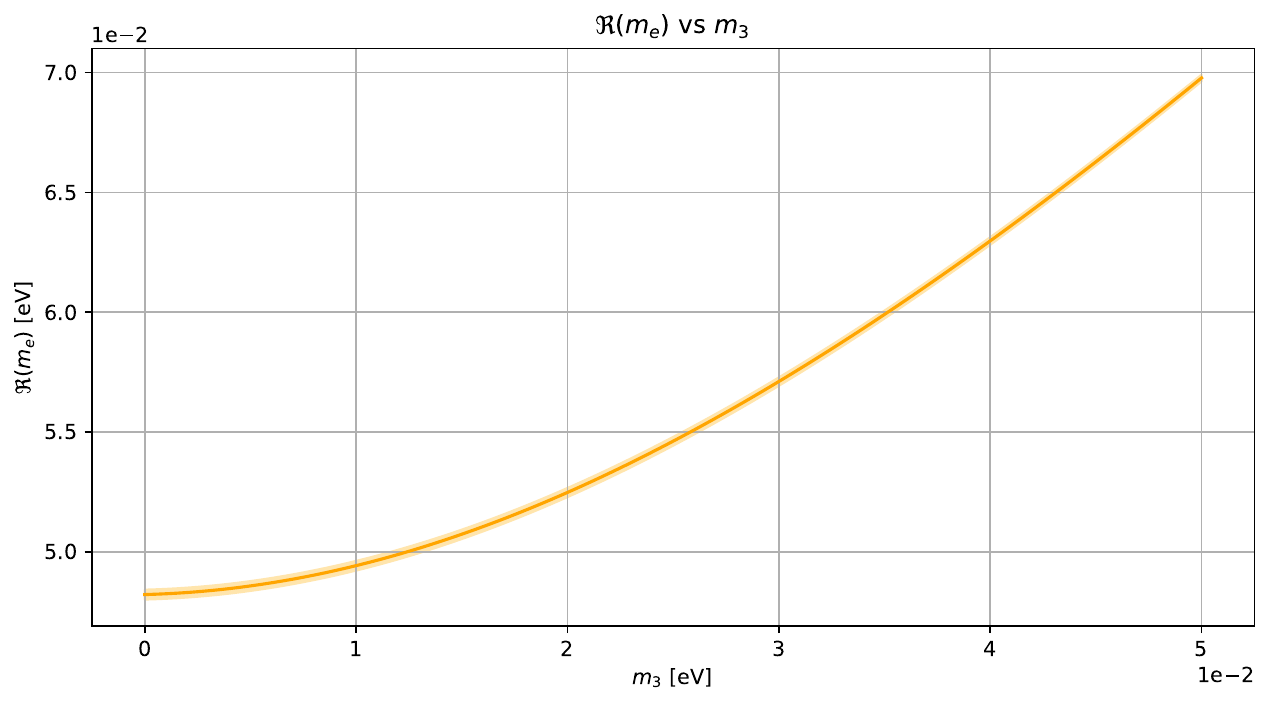}}
	\hspace{1cm}
	\subfigure
	{\label{fig:Re22_IH}
		\includegraphics[width=0.45\textwidth]{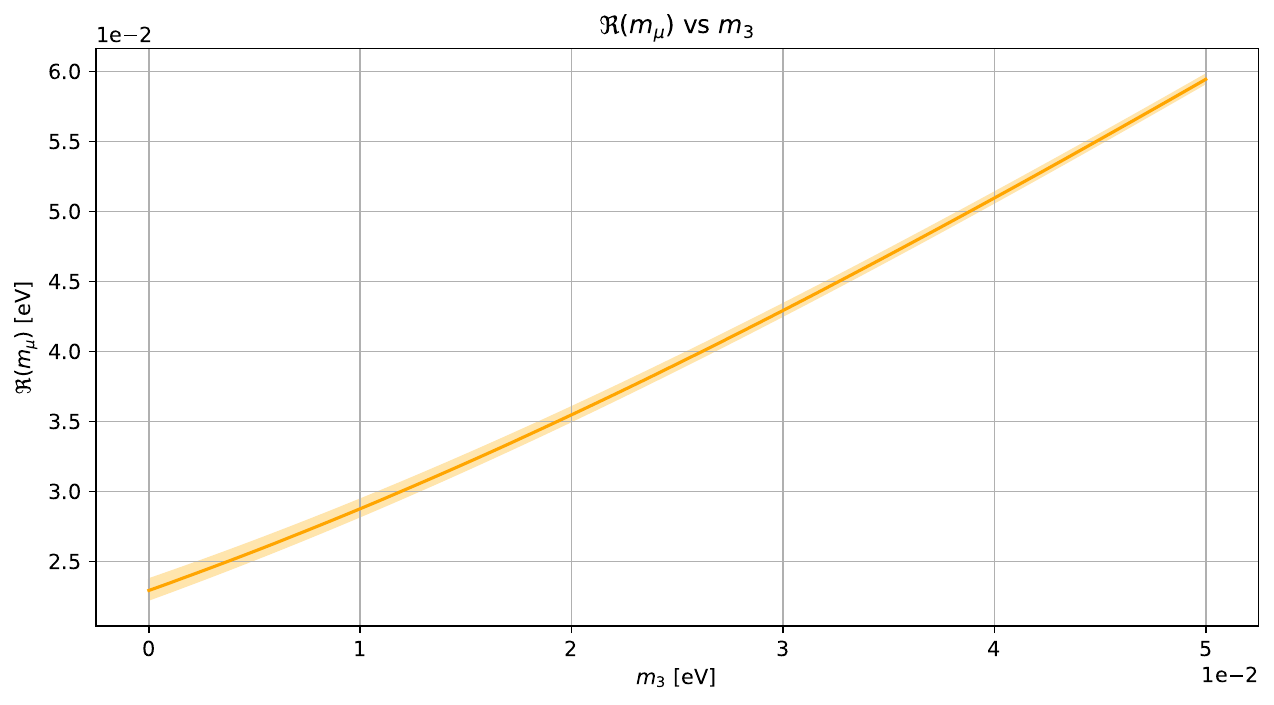}}
	\hspace{1cm}
	\subfigure
	{\label{fig:Re33_IH}
		\includegraphics[width=0.45\textwidth]{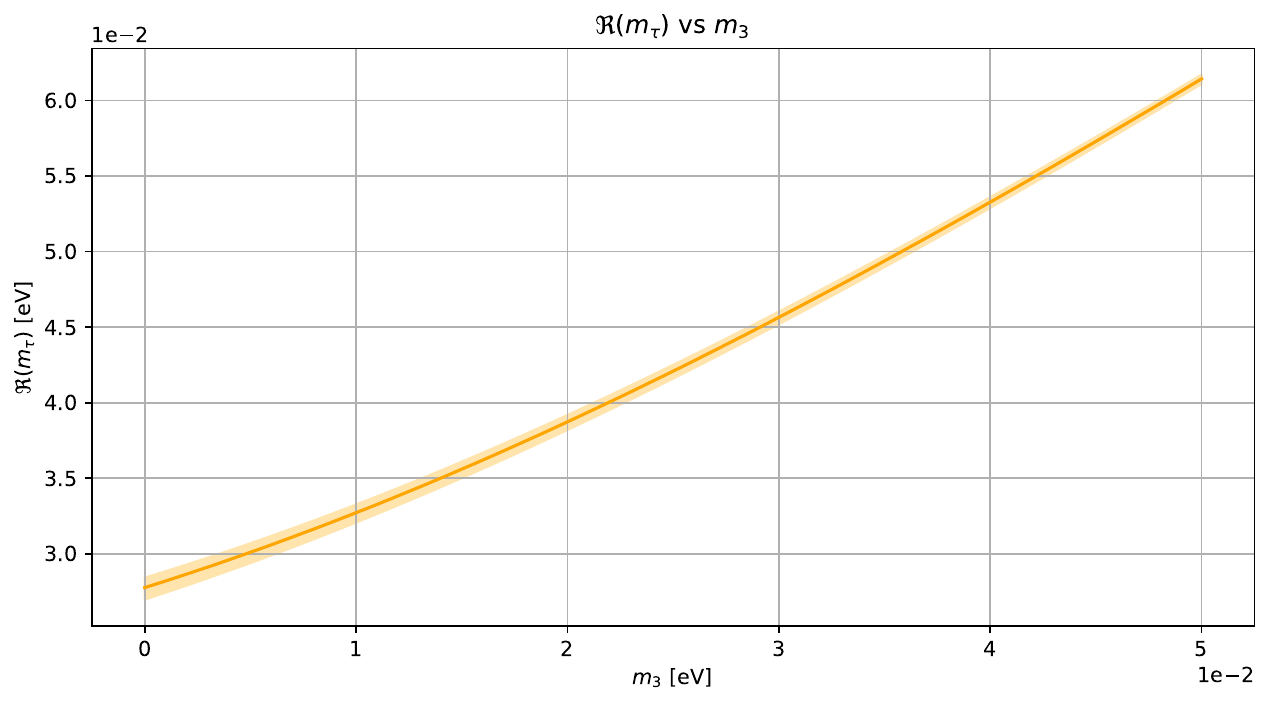}}
	\protect 
	\caption{Diagonal elements $m_{e}$, $m_{\mu}$, $m_{\tau}$ of the neutrino mass matrix in IH versus $m_3$}
\end{figure}
\begin{figure}[h!tbp]
	\centering
	\subfigure
	{\label{fig:Re12_IH}
		\includegraphics[width=0.45\textwidth]{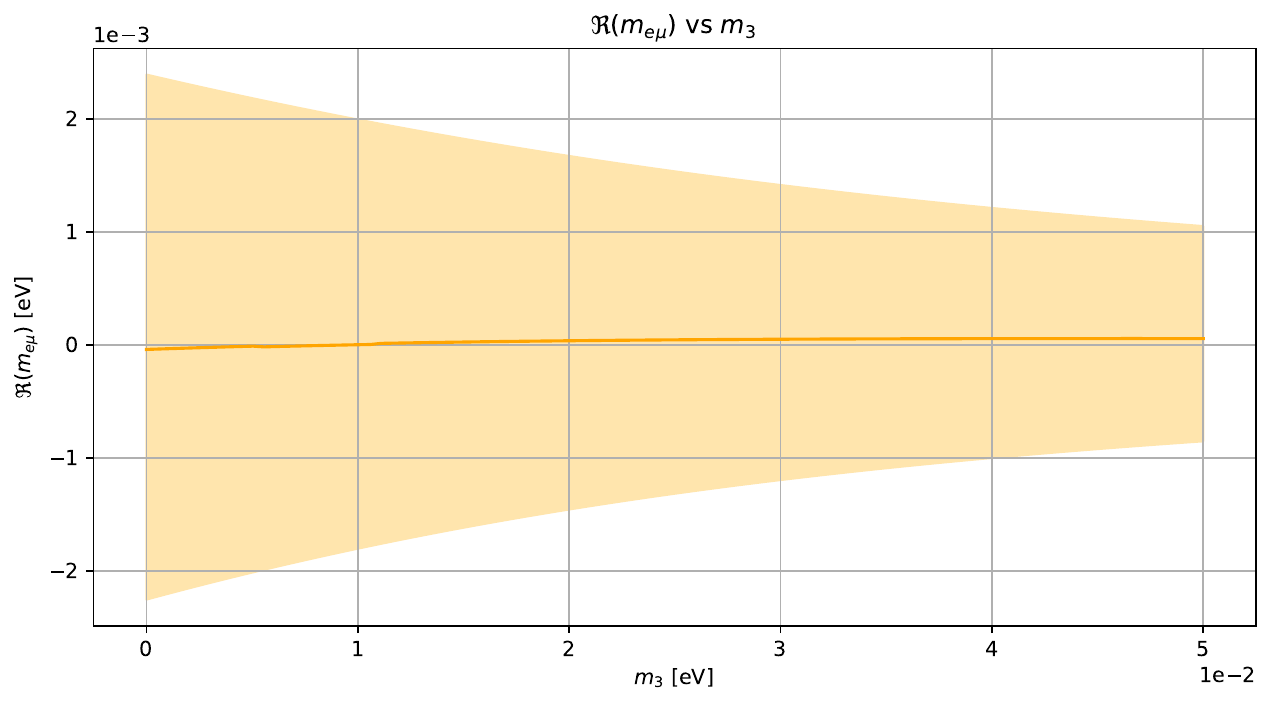}}
	\hspace{1cm}
	\subfigure
	{\label{fig:Im12_IH}
		\includegraphics[width=0.45\textwidth]{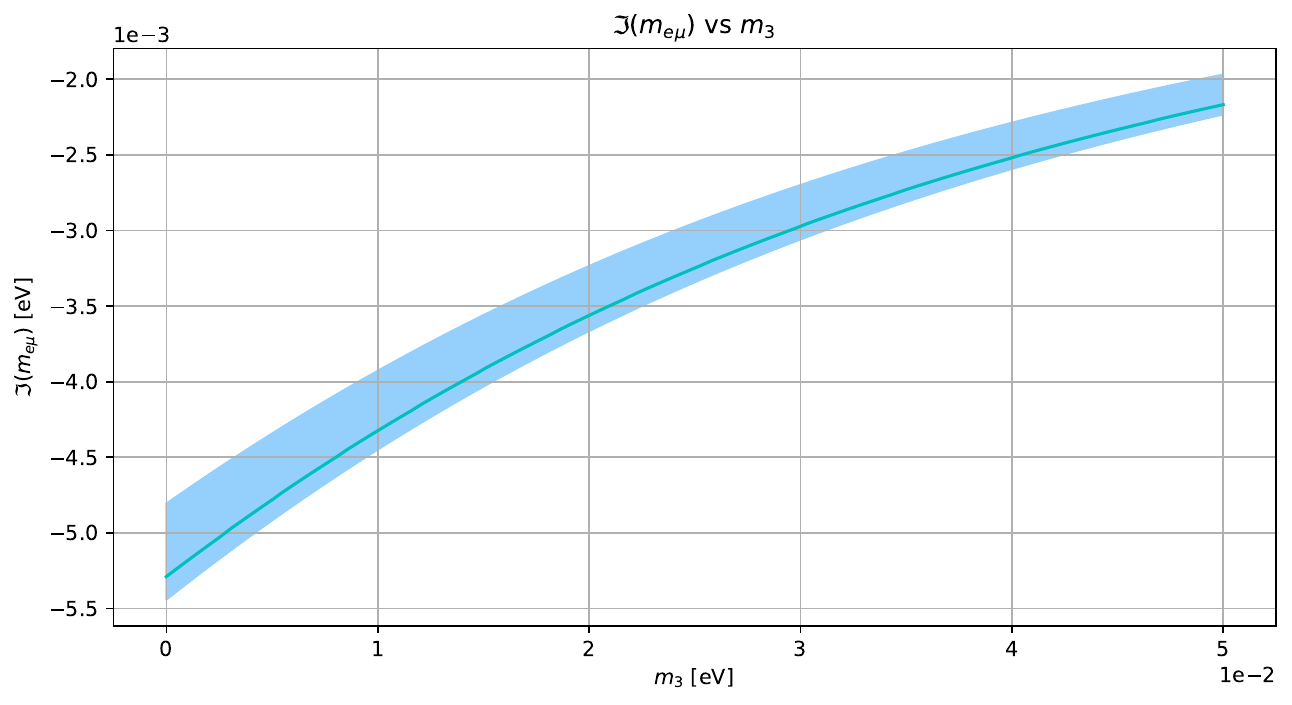}}
	\protect 
	\caption{Real and imaginary components of $m_{e\mu}$ in IH as a function of $m_3$}
\end{figure}
\begin{figure}[h!tbp]
	\centering
	\subfigure
	{\label{fig:Re13_IH}
		\includegraphics[width=0.45\textwidth]{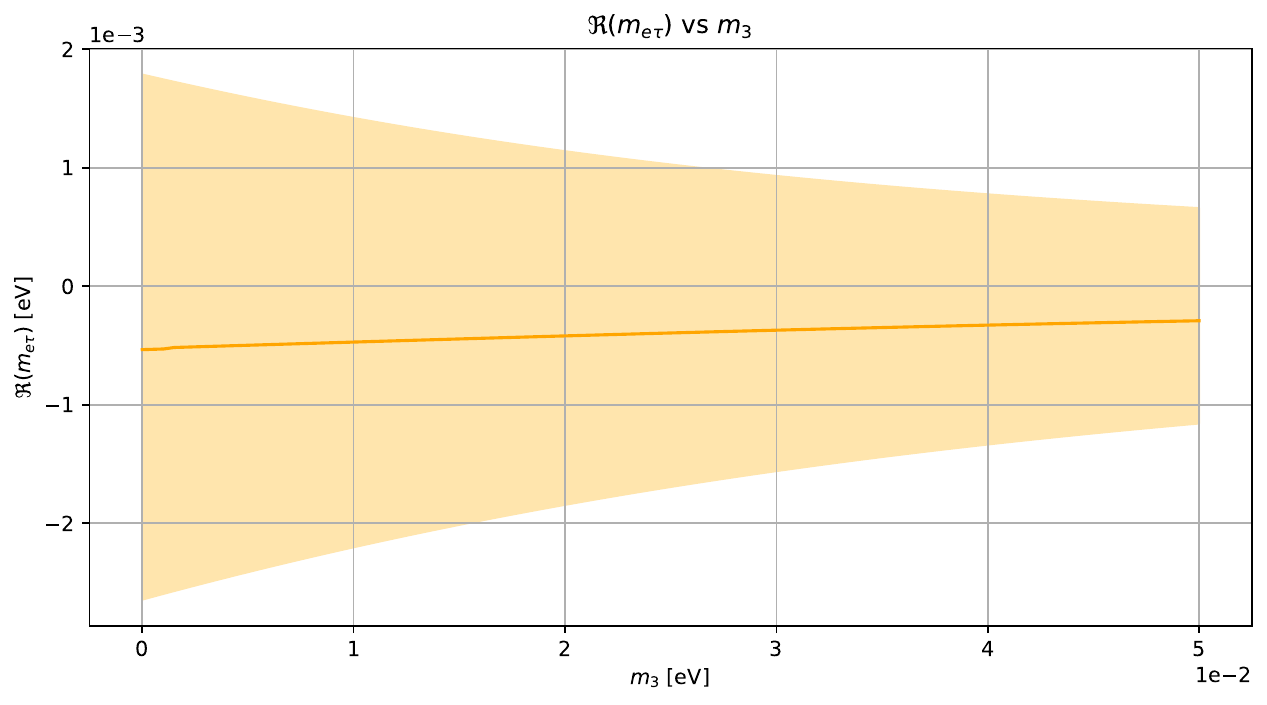}}
	\hspace{1cm}
	\subfigure
	{\label{fig:Im13_IH}
		\includegraphics[width=0.45\textwidth]{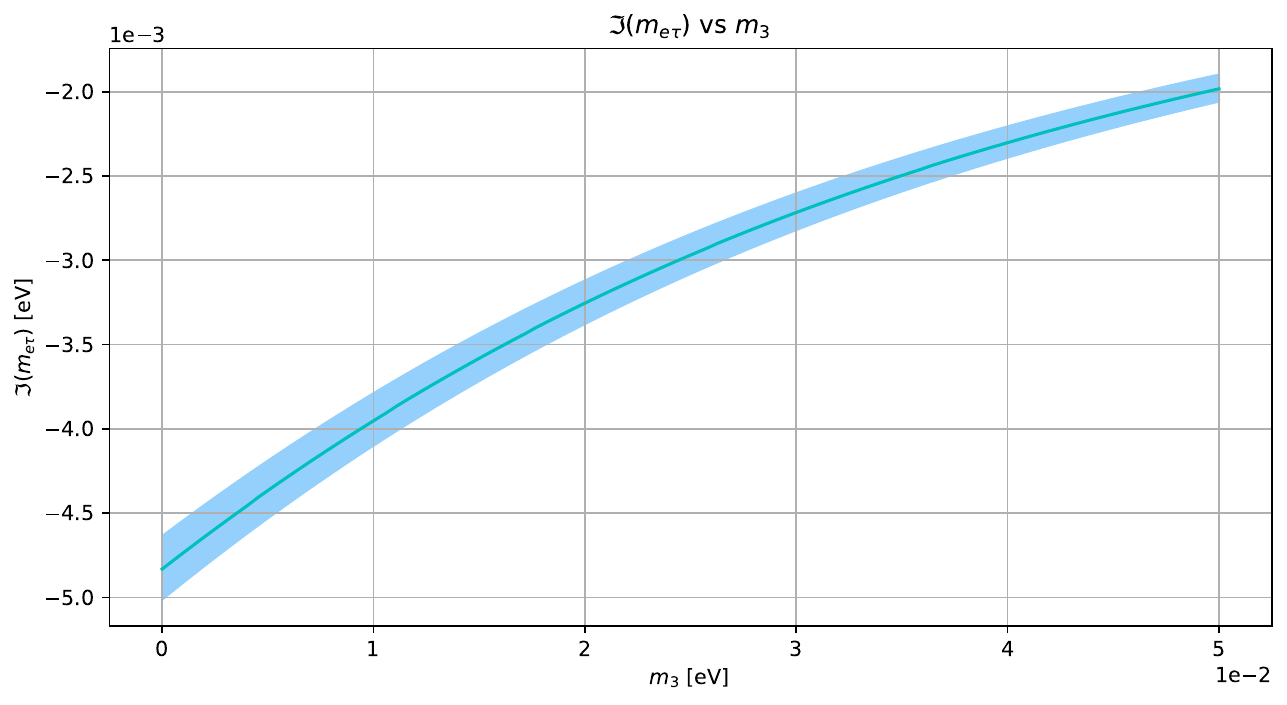}}
	\protect 
	\caption{Real and imaginary components of $m_{e\tau}$ in IH versus $m_3$}
\end{figure}
\begin{figure}[h!tbp]
	\centering
	\subfigure
	{\label{fig:Re23_IH}
		\includegraphics[width=0.45\textwidth]{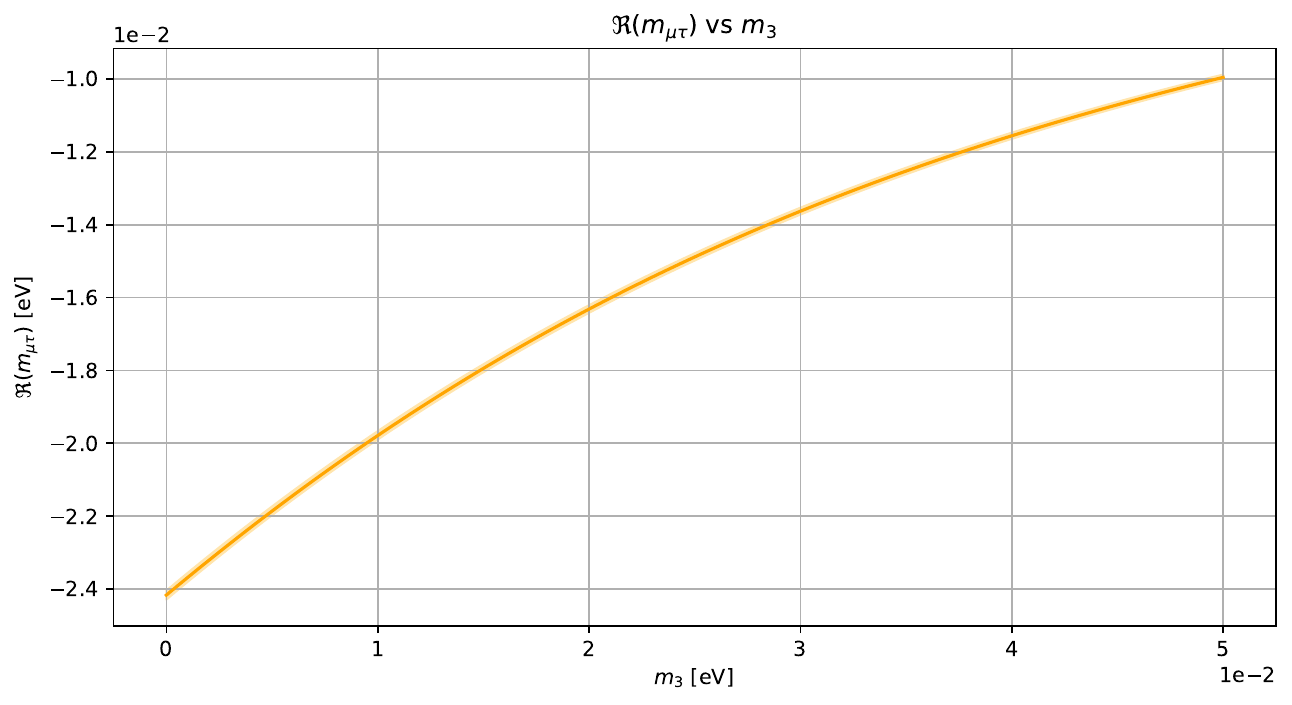}}
	\hspace{1cm}
	\subfigure
	{\label{fig:Im23_IH}
		\includegraphics[width=0.45\textwidth]{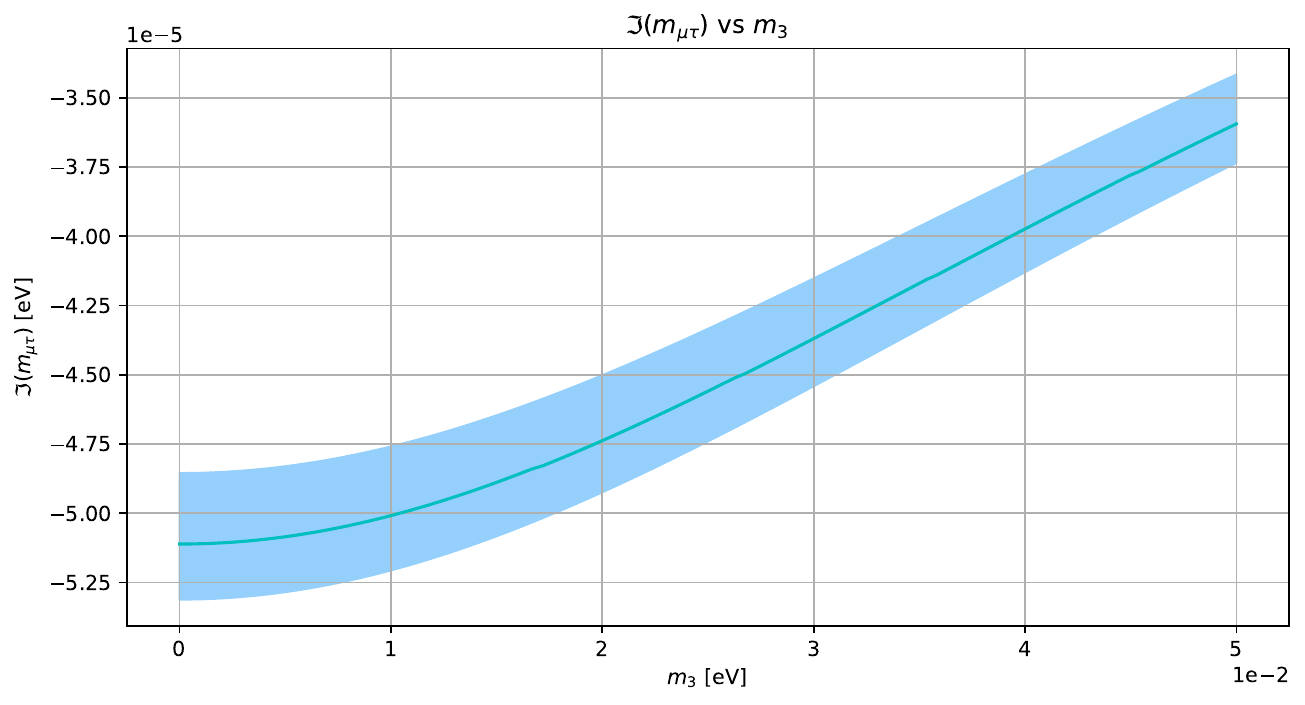}}
	\protect 
	\caption{Real and imaginary components of $m_{\mu\tau}$ in IH as a function of $m_3$}
\end{figure}

\end{document}